\documentclass[sn-mathphys-ay, iicol]{sn-jnl}

\usepackage{graphicx}%
\usepackage{multirow}%
\usepackage{amsmath,amssymb,amsfonts}%
\usepackage{amsthm}%
\usepackage{mathrsfs}%
\usepackage[title]{appendix}%
\usepackage{xcolor}%
\usepackage{textcomp}%
\usepackage{manyfoot}%
\usepackage{booktabs}%
\usepackage{algorithm}%
\usepackage{algorithmicx}%
\usepackage{algpseudocode}%
\usepackage{listings}%

\newtheorem{theorem}{Theorem}
\newtheorem{lemma}{Lemma}%
\newtheorem{example}{Example}%

\def\A{\mathbf A}
\def\B{\mathbf B}
\def\D{\mathbf D}
\def\H{\mathbf H}
\def\I{\mathbf I}
\def\M{\mathbf M}
\def\U{\mathbf U}

\def\X{\mathcal X}
\def\Y{\mathcal Y}
\def\Ac{\mathcal A}
\def\Nc{\mathcal N}
\def\supp{\mathcal S}
\def\Vc{\mathcal V}

\def\1{\mathbf 1}
\def\0{\mathbf 0}
\def\a{\mathbf a}
\def\b{\mathbf b}
\def\c{\mathbf c}
\def\g{\mathbf g}
\def\u{\mathbf u}
\def\t{\mathbf t}
\def\f{\mathbf f}
\def\s{\mathbf s}

\newcommand{\vv}{\mathbf v}
\newcommand{\w}{\mathbf w}
\newcommand{\x}{\mathbf x}
\newcommand{\y}{\mathbf y}
\newcommand{\z}{\mathbf z}

\def\NN{\mathbb N}
\def\PP{\mathbb P}
\def\RR{\mathbb R}
\def\SS{\mathbb S}
\def\TT{\mathbb T}

\def\tr{\mathrm{tr}}
\def\vech{\mathrm{vech}}
\newcommand{\rank}{\mathrm{rank}}

\begin{document}

\title[The Polytope of Optimal Approximate Designs]{The Polytope of Optimal Approximate Designs: Extending the Selection of Informative Experiments}


\author*[1]{\fnm{Radoslav} \sur{Harman}}\email{radoslav.harman@fmph.uniba.sk}
\author[1]{\fnm{Lenka} \sur{Filov\'{a}}}
\author[1]{\fnm{Samuel} \sur{Rosa}}

\affil*[1]{\orgdiv{Faculty of Mathematics, Physics and Informatics}, \orgname{Comenius University}, \orgaddress{\street{Mlynsk\'{a} Dolina F1}, \city{Bratislava}, \postcode{842 48}, \country{Slovakia}}}


\abstract{Consider the problem of constructing an experimental design, optimal for estimating parameters of a given statistical model with respect to a chosen criterion. To address this problem, the literature usually provides a single solution. Often, however, there exists a rich set of optimal designs, and the knowledge of this set can lead to substantially greater freedom to select an appropriate experiment. In this paper, we demonstrate that the set of all optimal approximate designs generally corresponds to a polytope. Particularly important elements of the polytope are its vertices, which we call vertex optimal designs. We prove that the vertex optimal designs possess unique properties, such as small supports, and outline strategies for how they can facilitate the construction of suitable experiments. Moreover, we show that for a variety of situations it is possible to construct the vertex optimal designs with the assistance of a computer, by employing error-free rational-arithmetic calculations. In such cases the vertex optimal designs are exact, often closely related to known combinatorial designs. Using this approach, we were able to determine the polytope of optimal designs for some of the most common multifactor regression models, thereby extending the choice of informative experiments for a large variety of applications.
}




\maketitle

\tableofcontents

\section{Introduction}

The optimal design of experiments is an important area of theoretical and applied statistics (e.g., \citet{Fedorov}, \citet{Pazman86}, \citet{Puk}, \citet{AtkinsonEA07}, \citet{Handbook}), closely related to various fields of mathematics and computer science, such as probability theory, orthogonal arrays, network theory, machine learning and mathematical programming (e.g., \citet{DetteStudden97}, \citet{Hedayat}, \citet{GhoshEA08}, \citet{He10}, \citet{Todd16}). In fact, one of the aims of this paper is to initiate a deeper exploration of the relationships between optimal experimental design, the theory of multidimensional polytopes, and rational-arithmetic computation.

From a statistical perspective, an optimal design is an experimental plan that maximizes the information contained in the observations.\footnote{In the following section, we will rigorously define the technical terms and prove all important claims used in the introduction.} The specific measure of information, called a criterion, depends on the goal of the experiment. The most prominent are the criteria of $D$- and $A$-optimality, both measuring the precision of the parameter estimation for the underlying statistical model.

We will primarily focus on ``approximate'' experimental design (also called designs ``for infinite-sample size'', or ``continuous'' designs), introduced by \citet{kiefer59}. An approximate design specifies a finite set of experimental conditions, which is called the support of the design, and the proportions of trials performed at each of these conditions, referred to as the design weights. In contrast, an exact design determines integer numbers of trials for the support points; often, these integer numbers are obtained from the proportions dictated by an approximate design, by means of appropriate rounding. We will use the term ``design'', without any qualifier, to denote specifically an approximate experimental design. 
\bigskip

The literature provides analytic forms of optimal designs (i.e., their supports and weights) for many statistical models and criteria. The focus is usually on obtaining a single optimal design; however, there is often an infinite set of optimal designs.

\begin{example}\label{ex:1}
Consider an experiment with $3$ continuous factors, the levels of which can be chosen anywhere within the interval $[-1,+1]$. Assume that for different trials, the observations are independent, and at any combination $(x_1,x_2,x_3)$ of factor levels, the observation satisfies the standard linear regression model 
\begin{equation}\label{eq:exam}
Y_{x_1,x_2,x_3}=\beta_1+\beta_2 x_1+\beta_3 x_2+\beta_4 x_3 + \epsilon,
\end{equation}
where $\beta_1,\beta_2,\beta_3,\beta_4$ are the unknown parameters, and $\epsilon$ is an unobservable error with $\mathrm{E}(\epsilon) \equiv 0$ and $\mathrm{Var}(\epsilon) \equiv \sigma^2 \in (0,\infty)$. Let us choose the criterion of $D$-optimality. Then, the uniform design that assigns $1/8$ of the trials to each of the $8$ factor level combinations in $\{-1,+1\}^3$, i.e., an approximate-design analogue of the full factorial design with $3$ binary factors, is optimal. However, consider any $\alpha \in [0,1]$ and a design that assigns $\alpha/4$ of the trials to each of the factor level combinations $(-1,-1,-1)$, $(+1,+1,-1)$, $(+1,-1,+1)$, $(-1,+1,+1)$, and $(1-\alpha)/4$ of the trials to each of the factor level combinations $(+1,+1,+1)$, $(-1,-1,+1)$, $(-1,+1,-1)$, $(+1,-1,-1)$. It is simple to show that such a design is also optimal. That is, there is a continuum of optimal designs.
\end{example}
\bigskip

The knowledge of more than one optimal design can be of theoretical as well as practical significance: alternative optimal designs may facilitate the best choice of the actual experiment. Of particular interest are optimal designs with small supports, because they may be suitable for assessing measurement errors, can be less expensive to execute, and can be more easily rounded to efficient exact designs. However, an optimal design with a large support may sometimes be more appropriate, because it may for instance better allow for the assessment of lack of fit of the model. Therefore, our purpose is to analyze the set of \emph{all} optimal designs for a given experimental design problem.
\bigskip

The available literature only provides sporadic results on the nonuniqueness of optimal design and on the selection of a particular optimal design. The fact that the optimal design is not always unique, and the problem of finding optimal designs with special characteristics, such as the ``minimal'' optimal designs, has already been mentioned by \citet{kiefer59} and \citet{FKW}. The term ``absolutely'' minimal optimal design was introduced by \citet{FKW} and later used by \citet{peso75}. Next, \citet{peso78} analyzed optimal designs for the quadratic regression on cubic regions and stated conditions for the uniqueness of the optimal design in this model.

In multifactor models, the simplest optimal design is often supported on a large product grid. For some models of this type, alternative optimal designs supported on smaller sets have been constructed (e.g., Corollary 3.5 in \citet{Cheng}, Theorem 2.1 in \citet{LimStudden88}, page 378 in \citet{Puk} and Section 5 in \citet{SchwabeWong}). A result of \citet{SchwabeWierich} provides conditions that characterize all $D$-optimal designs in additive multifactor models with constant term and without interactions via the $D$-optimal designs in the single-factor models. Further, for the additive model in crossover designs, there can be a multitude of optimal designs, and \citet{Kushner} suggests searching for them using linear programming. Focusing on applications, \citet{robinson} analyze the situation when there are several exact $D$-optimal screening designs and suggest choosing the suitable one subject to a secondary characteristic. More recently \citet{RosaHarman16} employed linear programming to select optimal designs with small supports in the models with treatment and nuisance effects. 
\bigskip

In this paper, unlike most of the optimal design literature, we do not solve the problem of constructing a single optimal design. Instead, we assume that we already have one, or that it can be easily constructed using theoretical or numerical methods. We demonstrate that for a given model and criterion, in cases where the optimal design is not unique, the set of all optimal designs is typically a polytope. Additionally, we provide various theoretical properties of this polytope. We show that its vertices, which we call vertex optimal designs, can not only be used to succinctly describe all other optimal designs, but themselves have various useful characteristics, such as small supports.

\begin{example}\label{ex:2}
In Example \ref{ex:1}, the uniform design on $\{-1,+1\}^3$ is a trivial $D$-optimal design. However, the set of designs from Example \ref{ex:1} parametrized by $\alpha \in [0,1]$ is exactly the set of \emph{all} $D$-optimal designs for Model \eqref{eq:exam}. That is, the polytope of optimal designs is a line segment, with two vertices. These vertices correspond to the two vertex optimal designs, defined by $\alpha=0$ and $\alpha=1$; they are the approximate design analogues of the two $D$-optimal $2^{3-1}$ fractional factorial designs. The vertex optimal designs are the only minimal optimal designs. That is, they are the only optimal designs with the following property: in the sense of set inclusion, there is no optimal design with a smaller support. On the other hand, note that the support of any optimal design corresponding to $\alpha \in (0,1)$ is \emph{maximal} in the sense that it covers the supports of all optimal designs.
\end{example} 
\bigskip

As we will show, the polytope of optimal designs can be described in a relatively straightforward way as an intersection of halfspaces, which is called the H-representation of a polytope. However, enumerating the vertices of a polytope given by its H-representation generally requires extensive calculations. This is known as the vertex enumeration problem.

The vertex enumeration problem can be solved by the pivoting algorithms based on the simplex method of linear programming (e.g., \citet{Avis-Fukuda}) and the nonpivoting algorithms based on the double description method (\citet{DD}, \citet{avis-jordan}). An overview of the vertex enumeration methods can be found in \citet{Matheiss} or Chapter 8 in \citet{fukuda}. In experimental design, the double description algorithm has been used by \citet{Snee} and by \citet{CoetzerHaines} to compute the extreme points of the design space in Scheff\`e mixture models with multiple linear constraints. In \citet{He}, the algorithm facilitated the construction of a special type of minimax space-filling designs.

Computationally, the general vertex enumeration problem is NP-complete (\citet{Khachiyan}). Despite its theoretical complexity, a key property of this problem is that if the H-representation of the polytope involves only rational numbers, the coordinates of its vertices are also rational numbers. Moreover, in this case the vertex enumeration algorithms can execute all calculations via rational arithmetic, thus providing an error-free method of constructing the vertices (e.g., \citet{cdd}), which is akin to a computer-assisted proof. For a general exposition of error-free computing in geometric problems, see \citet{Eberly}.

This fact has important consequences for the construction of vertex optimal designs of experiments. The reason is that for many optimal design problems, the H-representation of the optimal polytope is fully rational. Consequently, by employing the rational-arithmetic computations, we can provide a complete list of all vertex optimal designs for a variety of standard models and a large class of criteria, involving both $D$- and $A$-optimality. In addition, for the rational optimal design problems, the vertex optimal designs are exact, i.e., realizable with a finite number of trials. Often, they correspond to standard combinatorial designs of experiments.

It is also possible to apply floating-point versions of the vertex enumeration algorithms for arbitrary optimal design problems as a general algorithmic method of constructing the vertex optimal designs. However, the outputs of such algorithms are not always perfectly precise due to the nature of floating-point computations. Since our aim is to provide results that are completely reliable from the theoretical point of view, we only focus on rational optimal design problems.
\bigskip

The rest of this paper is organized as follows. Section \ref{subsec:ODs} sets the background on the theory of optimal design. In Section \ref{subsec:mos}, we introduce the notions of a maximal optimal design and maximum optimal support, which play important roles in the construction of the polytope of optimal designs. In Sections \ref{subsec:PODW} and \ref{subsec:POD}, we mathematically describe the optimal polytopes viewed as sets of finite-dimensional vectors and as finitely supported probability measures, respectively. Section \ref{subsec:Rational} introduces the class of rational optimal design problems, for which it is possible to completely and reliably determine the optimal polytopes. Section \ref{subsec:Specific} provides suggestions for utilizing the polytope of optimal designs for constructing experimental designs with specific properties. In Section \ref{sec:Examples}, we characterize the polytopes of optimal designs for various common models and criteria.  Section \ref{sec:Further} offers concluding remarks, highlighting key insights of our findings and avenues for further study. Additionally, appendices include a reference list of the notation, all non-trivial proofs, and other supporting information.

\section{Theory}\label{sec:Theory}

\subsection{Optimal design}\label{subsec:ODs}

Let $\X \subset \RR^k$ be a compact design space, that is, the set of all design points available for the trials. Note that such a definition of $\X$ includes both continuous and discrete spaces. 

Consider a continuous function $\f: \X \to \RR^m$, $m \geq 1$, that defines a linear regression model on $\X$ in the following sense: provided that the experimental conditions of a trial are determined by the design point $\x \in \X$, the observation is $Y_\x = \f'(\x)\beta + \epsilon$, where $\beta \in \RR^m$ is a vector of unknown parameters, and $\epsilon$ is an unobservable random error with expected value $0$ and variance $\sigma^2 \in (0, \infty)$. The random errors across different trials are assumed to be independent.

We remark that we restrict ourselves to linear homoscedastic regression models only for simplicity; the results of this paper can also be applied to nonlinear regression models through the approach of local optimality and to cases of non-constant error variance by means of a simple model transformation (e.g., \cite{AtkinsonEA07}, Chapters 17 and 22).
\bigskip

As is usual in the optimal design literature (e.g., Section 1.24 in \cite{Puk}), we will formalize an approximate design as a probability measure $\xi$ on $\X$ with a finite support $\supp(\xi):=\{\x_1,\ldots,\x_h\}$, where $h$ is the support size, and $\x_1,\ldots,\x_h \in \X$ are the support points. The interpretation is that for any $\x \in \supp(\xi)$ the number $\xi(\x):=\xi(\{\x\})>0$ gives the proportion of the trials to be performed under the experimental conditions determined by the design point $\x$. For clarity, we will sometimes represent the design $\xi$ by the table
\begin{equation*}
\left[
\begin{array}{cccc}
\x_1 & \x_2 & \cdots & \x_h \\ \hline
\xi(\x_1) & \xi(\x_2) & \cdots & \xi(\x_h) \\
\end{array}
\right].
\end{equation*}

For a design $\xi$, the normalized information matrix is the positive semi-definite matrix
\begin{equation*}
    \M(\xi)=\sum_{\x \in \supp(\xi)}\xi(\x)\f(\x)\f'(\x),
\end{equation*}
which is of size $m \times m$. We will assume that the problem at hand is nondegenerate in the sense that there is a design $\xi$ such that $\M(\xi)$ is nonsingular. 
\bigskip

The quality of the design $\xi$ is measured by the ``size'' of $\M(\xi)$, assessed through a utility function known as a criterion. For simplicity,\footnote{It is straightforward to use the methods developed in this paper with any information function $\Phi$ that is strictly concave in the sense of Section 5.2 in \cite{Puk}.} we will focus on Kiefer's criteria $\Phi_p$ with a parameter $p \in (-\infty, 0]$ in their information-function versions, as defined in Chapter 6 of \citet{Puk}. For a singular $\M(\xi)$, the value of the $\Phi_p$-criterion is $0$, and for a non-singular $\M(\xi)$ it is defined by 
\begin{equation*}
\Phi_p(\M(\xi)) = 
\begin{cases}
    \det^{1/m}(\M(\xi)), & \text{if } p = 0, \\
    m^{-1/p}\tr^{1/p}(\M^p(\xi)), & \text{if } p < 0.
\end{cases}
\end{equation*}

This class includes the two most important criteria: $D$-optimality ($p=0$) and $A$-optimality ($p=-1$). We also implicitly cover the criterion of $I$-optimality (also called $V$- or $IV$-optimality), because the problem of $I$-optimal design is equivalent to the problem of $A$-optimal design in a linearly transformed model (e.g., Section 9.8 in \citet{Puk}).

For the chosen $p$, the $\Phi_p$-optimal design is any design $\xi^*$ that maximizes $\Phi_p(\M(\xi))$ in the set of all designs $\xi$. Note that our assumptions imply that there exists at least one $\Phi_p$-optimal design.

The most important result on optimal approximate designs is the so-called equivalence theorem; for Kiefer's criteria it states that a design $\xi^*$ is $\Phi_p$-optimal if and only if $\M(\xi^*)$ is non-singular and
\begin{equation}\label{eq:eqt21}
    \f'(\x)\M^{p-1}(\xi^*)\f(\x) \leq \tr(\M^p(\xi^*)) \text{ for all } \x \in \X,
\end{equation}
with equality for all $\x \in \supp(\xi^*)$ (e.g., Section 7.20 in \citet{Puk}). In addition, for several standard linear regression models (e.g., those analyzed in Sections \ref{ssMod2}, \ref{subsec:main} and \ref{ss:Mod4}), there exists a simple design $\xi^*$ such that $\M(\xi^*)=\lambda \I_m$ for some $\lambda>0$ and
\begin{equation}\label{eq:eqtSchur}
    \max_{\x \in \X} \|f(\x)\|^2 = m\lambda.
\end{equation}
Then, by Corollary 4 of \cite{Harman08}, the design $\xi^*$ is $\Phi_p$-optimal for all parameters $p$.
\bigskip

Let $p$ be fixed. While there can be an infinite number of $\Phi_p$-optimal designs, the properties of the Kiefer's criteria and the assumption of nondegeneracy imply that the information matrix of all $\Phi_p$-optimal designs is the same; see, e.g., Section 8.14 in \citet{Puk}. We will denote this unique and necessarily nonsingular $\Phi_p$-optimal information matrix by $\M_*$. In the rest of this paper, if the criterion is general, fixed, or known from the context, we will drop the prefix $\Phi_p$-.
\bigskip

Of particular importance are the notions of a minimal optimal design and an absolutely minimal optimal design, introduced by \citet{FKW}. An optimal design $\xi^*$ is a minimal optimal design if there is no optimal design, whose support is a proper subset of $\supp(\xi^*)$.\footnote{Some authors use the expression ``minimally-supported design'' to denote any design with the support size equal to the number of parameters (\citet{ChangLin}, \citet{LiMajumdar} and many others), which is a different concept.} An optimal design $\xi^*$ is called an absolutely minimal optimal design if there is no optimal design, whose support has fewer elements than $\supp(\xi^*)$. 

\subsection{Maximum optimal support}\label{subsec:mos}

As stated in the introduction, our aim is to characterize the set of all optimal designs, if we already know one optimal design. To this end, we will assume that we know a ``maximal'' optimal design $\xi^*$, which we define as an optimal design satisfying $\supp(\xi^*) \supseteq \supp(\tilde{\xi}^*)$ for any other optimal design $\tilde{\xi}^*$. We emphasize that in many models, such as those analysed in Section \ref{sec:Examples}, the known and simplest optimal design $\xi^*$ satisfies
\begin{equation}\label{eq:eqt22}
    \f'(\z)\M^{p-1}_*\f(\z) < \tr(\M^p_*)\text{ for all }\z \in \X \setminus \supp(\xi^*),
\end{equation}
i.e, $\xi^*$ is maximal, owing to the equivalence theorem for $\Phi_p$-optimality. 

Of course, for some optimal design problems, a maximal optimal design is not immediately apparent. Even then, it is usually possible to construct one by a procedure that inspects either the design points satisfying equality in \eqref{eq:eqt21}, or the points of a theoretically derived finite set necessarily covering the supports of all optimal designs; see, e.g., Section 3 in \cite{Heil92} or Lemma 4.2 in \cite{GGS14}.\footnote{Since there is a large variety of models with directly available maximal optimal designs to justify the broad applicability of our method, see Section \ref{sec:Examples}, we decided not to detail such an inspection procedure or the theoretical results on possible supports of optimal designs} in this paper. Note that some very specific models, such as the trigonometric regression on $[0, 2\pi]$ (Section 9.16 in \citet{Puk}), do not admit any maximal optimal design. This is because in these models the union of the supports of all optimal designs is an infinite set and, according to the standard definition we use, any design only has a finite support. We exclude such optimal design problems from our further considerations.
\bigskip

While there can be many maximal optimal designs, their supports must be the same; we will denote the unique ``maximum optimal support'' by $\Y:=\{\y_1,\ldots,\y_d\}$, where $d$ is the size of the maximum optimal support, and the regressor vectors corresponding to the trials at $\Y$ by $\f_i:=\f(\y_i)$, for all $i=1,\ldots,d$.

Therefore, if our aim is to study only optimal designs, we can completely restrict our attention to the set $\Y$ and disregard the design points from $\X \setminus \Y$. Let us henceforth denote by $\Upsilon$ the set of all designs supported on $\Y$. 

\begin{example}\label{ex:3}
Consider Model \eqref{eq:exam} from Examples \ref{ex:1} and \ref{ex:2}. Here, the design space is $\X=[-1,+1]^3$; that is, we have $k=3$ continuous factors. The regression function is $\f(\x)=(1,x_1,x_2,x_3)'$ for all $\x=(x_1,x_2,x_3)' \in \X$, i.e., the number of parameters is $m=4$. Let $\xi^*$ be the uniform design on the set $\Y=\{-1,+1\}^3$ with $d=8$ elements. A simple calculation yields $\M(\xi^*)=\sum_{\y \in \Y} \frac{1}{8}\f(\y)\f'(\y)=\I_4$. The design $\xi^*$ is optimal for any $\Phi_p$, because $\|\f(\x)\|^2 \leq 4$ for all $\x \in \X$, which is equivalent to \eqref{eq:eqt21}. In addition, we clearly have $\|\f(\z)\|^2 < 4$ for all $\z \in \X \setminus \Y$, which is a transcription of \eqref{eq:eqt22}; hence, $\xi^*$ is a maximal optimal design, and $\Y$ is the maximum optimal support. All $\Phi_p$-optimal designs for Model \eqref{eq:exam} therefore belong to the set $\Upsilon$ of designs supported on $\Y$.
\end{example}

\subsection{Polytope of optimal weights}\label{subsec:PODW}

Because $\Y$ is finite, any design $\zeta$ in $\Upsilon$ can be characterized by the weights that it assigns to the design points $\y_1, \ldots, \y_d$. We will call $\w_\zeta = (\zeta(\y_1),\ldots,\zeta(\y_d))'$ the vector of weights of $\zeta$. This correspondence allows us to work with vectors from $\RR^d$ instead of measures on $\Y$. The components of each $\w_\zeta$ must be non-negative and sum to one; the set of all weight-vectors on $\Y$ is therefore the probability simplex $\PP := \{\w \in \RR^d : \1'_d\w=1, \w \geq \0_d\}$. Analogously, given a vector of weights $\w \in \PP$, the design $\zeta_\w$ determined by $\w$   satisfies $\zeta_\w(\y_i) = w_i$ for all $i = 1, \ldots, d$. 
\bigskip

Because all optimal designs have the same information matrix, the set of all weight-vectors corresponding to the optimal designs is 
\begin{equation*}
    \PP_* := \{\w \in \PP : \M(\zeta_\w)=\M_*\},
\end{equation*}
where $\M(\zeta_\w)=\sum_{i=1}^d w_i \f_i \f'_i$. Since $\PP$ is a bounded superset of $\PP_*$, and the equalities $\M(\zeta_\w)=\M_*$ are linear in $\w$, the set $\PP_*$ is a polytope\footnote{Because the definition of the term ``polytope'' somewhat varies across the mathematical literature, we remark that by polytope we mean a bounded convex polyhedron, i.e., a bounded intersection of closed half-spaces. For an introduction to the theory of polytopes, see, e.g., \citet{ziegler}.} in $\RR^d$. Therefore, we will say that $\PP_*$ is the ``polytope of optimal weights''. A compact expression of the polytope of optimal weights utilizes the matrix 
\begin{equation*}
    \A:=\bigl[\vech\left(\f_1\f'_1\right), \ldots, \vech\left(\f_d\f'_d\right)\bigr],
\end{equation*}
where $\vech$ means the half-vectorization of a symmetric matrix. Therefore, the number of rows of A is $q:=\frac{1}{2}m(m+1)$.

\begin{lemma}\label{lPolyShort} The set of optimal design weights is
    \begin{equation}\label{eq:H}
       \PP_*=\{\w \geq \0_d: \A \w = \vech(\M_*)\}.
    \end{equation}
\end{lemma}
The proof of Lemma \ref{lPolyShort} requires showing that $\M(\zeta_\w)=\M_*$ for some $\w \geq 0$ entails $\1_d'\w=1$; see Appendix \ref{sup:proofs} for the proofs. Note that \eqref{eq:H} is geometrically the intersection of a finite number of halfspaces, that is, it is a H-representation of $\PP_*$.  
\bigskip

For the mathematical properties of the optimal design problem at hand, a key characteristic is the rank $s$ of $\A$, which we will call the ``problem rank''. In the following lemma, ``dimension'' of a subset of a vector space refers to the affine dimension of its affine hull. 
\begin{lemma}\label{lMsm}
    (a) The bounds for the problem rank are $m \leq s \leq \min(d,q)$; (b) The dimension of the set $\{\M(\zeta): \zeta \in \Upsilon\}$ of the information matrices of all designs supported on $\Upsilon$ is $s-1$; (c) The dimension of the polytope $\PP_*$ of optimal design weights is $t:=d-s$. 
\end{lemma}

A simple consequence of Lemma \ref{lMsm} is that the polytope $\PP_*$ is trivial ($t=0$), i.e., the optimal design is unique, if and only if $d=s$, which is if and only if $\A$ has full column rank. Note also that Lemma \ref{lMsm}, parts (a) and (c), imply $t \leq d-m$. It follows that for the case $d=m$ the polytope $\PP_*$ is trivial.  
\bigskip

Because the dimension of $\PP_*$ is $t < d$, $\PP_*$ belongs to some proper affine subspace of $\RR^d$. However, we can transform $\PP_*$ to a full-dimensional polytope $\TT^* \subset \RR^t$ that has the same geometric properties (the number and incidence of the faces of all dimensions, the distance between vertices and so on), as follows.

Let $t>0$, let the columns of $\U=(\u_1,\ldots,\u_t)$ form an orthonormal basis of the null space $\Nc(\A)$ and let $\xi^*$ be a maximal optimal design with the vector of weights $\w^* > \0_d$. Define an affine mapping $A$ at $\RR^t$ by $A(\t)= \U\t+\w^*$. Note that the image of $A$ is the affine space $\Ac=\Nc(\A) + \w^* \supset \PP_*$, and its inverse on $\Ac$ is $A^{-1}(\w)=\U'(\w - \w^*)$. Therefore, $A$ is an affine bijection between $\RR^t$ and $\Ac$, which means that the geometric structure of $\PP_*$ is the same as the structure of
\begin{eqnarray}\label{eq:isomorph}
    \TT_*:=A^{-1}(\PP_*)=\{\t \in \RR^t : \U\t \geq - \w^*\}.
\end{eqnarray}
The interior of $\TT_*$ contains the origin $\0_t$. The full-dimensional polytope $\TT_*$ can be used for the numerical construction of specific optimal designs as outlined in Section \ref{subsec:Specific}, as well as for a visualization of the set of all optimal designs if $t \leq 3$.
\bigskip

Key elements of $\PP_*$ are its vertices. Note that the number $\ell$ of these vertices is finite; we will denote them $\vv_1,\ldots,\vv_\ell$. The representation \eqref{eq:H} of $\PP_*$ is in fact in the standard form of the feasible set of a linear program. As such, the well-established theory of linear programming can be used to give the following characterizations of $\vv_1,\ldots,\vv_\ell$.

\begin{theorem}\label{tEquivPOW}
For $\w \in \RR^d$ let $\supp(\w)=\{i \in \{1,\ldots,d\} : w_i \neq 0\}$. The following statements are equivalent:
\begin{enumerate}
    \item $\vv$ is a vertex of $\PP_*$, i.e., $\vv$ is the  unique solution to the optimization problem $\min \{\c'\w: \w \in \PP_*\}$ for some $\c \in \RR^d$.
    \item $\vv$ is an extreme point of $\PP_*$, i.e., $\vv \in \PP_*$, and if $\w_1,\w_2 \in \PP_*$ are such that $\vv = \alpha \w_1 + (1-\alpha)\w_2$ for $\alpha \in (0,1)$ then $\w_1=\w_2=\vv$. 
    \item $\vv \in \PP_*$, and the vectors $\{\vech\left(\f_i\f'_i\right): i \in \supp(\vv)\}$ are linearly independent.
    \item $\vv \in \PP_*$, and if $\w \in \PP_*$ is such that $\supp(\w) \subseteq \supp(\vv)$, then $\supp(\w) = \supp(\vv)$.
    \item $\vv \in \PP_*$, and if $\w \in \PP_*$ is such that $\supp(\w) \subseteq \supp(\vv)$, then $\w = \vv$.
\end{enumerate}
\end{theorem}
\begin{example}\label{ex:4}
Consider Model \eqref{eq:exam} from Examples \ref{ex:1}-\ref{ex:3}, with the maximum optimal support $\Y = \{-1,+1\}^3$ and the optimal information matrix $\M_*=\I_4$. Denote the elements of $\Y$ lexicographically as $\y_1 = (-1,-1,-1)', \y_2 = (-1,-1,+1)', \ldots, \y_8 = (+1,+1,+1)'$. By Lemma \ref{lPolyShort}, $\PP_* = \{ \w \in \RR^8 : \A\w = \vech(\I_4), \w \geq \0_8 \}$, where the columns of $\A$ are $\vech((1,\y'_i)'(1,\y'_i))$, $i=1,\ldots,8$. The full form of $\A\w = \vech(\I_4)$ is provided as equation \eqref{eq:Awb} in Appendix \ref{sup:matrix}. It is a matter of simple algebra to verify that the problem rank is $\rank(\A)=s=7$. Consequently, part (c) of Lemma \ref{lMsm} implies that the dimension of $\PP_*$ is $t=1$. Laborious but straightforward calculations show that all solutions to the system $\A\w = \vech(\I_4)$ are $\w = \1_8/8 + c(\s_1 - \s_2)$, where $c \in  \RR$, $\s_1= (0,1,1,0,1,0,0,1)'$, and $\s_2= (1,0,0,1,0,1,1,0)'$. Adding $\w \geq \0_8$ restricts the values of $c$ to the interval $[-1/8, 1/8]$. Therefore, the elements of $\PP_*$ can be parametrized as $\w_\alpha = \s_2/4 + (\alpha/4)(\s_1 - \s_2)$, $\alpha \in [0,1]$. Hence, $\PP_*$ is the line segment in $\RR^8$ with endpoints (vertices) $\vv_1=\s_1/4$ and $\vv_2=\s_2/4$. Finally, since the vector of weights of a maximal optimal design is $\w^* = \frac{1}{8}\1_8$, $\Nc(\A)$ has dimension $t=1$, and its one-element orthonormal basis is $\U = (\s_1-\s_2)/\sqrt{8}$, formula \eqref{eq:isomorph} yields that a full-dimensional ``polytope'' isomorphic to $\PP_*$ is the interval $\TT_* = [-\sqrt{8}, \sqrt{8}]$. For this simple problem, one can also directly verify the characterization of the vertices $\vv_1$ and $\vv_2$ given by Theorem \ref{tEquivPOW}.
\end{example}

\subsection{Polytope of optimal designs}\label{subsec:POD}

The central object of this paper is the set $\Upsilon_*$ of all $\Phi_p$-optimal designs, i.e.,
\begin{equation*}
    \Upsilon_*:=\{\zeta \in \Upsilon: \M(\zeta)=\M_*\}.
\end{equation*}
Here we formulate corollaries of the results of Section \ref{subsec:PODW} in the standard optimal design terminology, so that it is clear how these results translate from the weights $\w \in \RR^d$ to actual designs (i.e., finitely supported probability measures).

First, we show that the mathematical properties indeed translate to designs $\zeta \in \Upsilon$ by formalizing the transformation $\zeta \mapsto \w_\zeta$ as a linear isomorphism. Let $\Upsilon^{\pm}$ be the linear space of all signed measures supported on $\Y$. Clearly, $L: \zeta \to \w_\zeta=(\zeta(\y_1),\ldots,\zeta(\y_d))'$ is a linear one-to-one mapping between $\Upsilon^{\pm}$ and $\RR^d$, that is, a linear isomorphism, with inverse $L^{-1}:\w \to \zeta_\w$. Note that the image of $\Upsilon$ in $L$ is $\PP$, and the image of $\Upsilon_*$ in $L$ is $\PP_*$. The latter observation means that $\Upsilon_*$ is isomorphic to $\PP_*$; we therefore call $\Upsilon_*$ the ``polytope of optimal designs''. 
\bigskip

The Carath\'{e}odory theorem (e.g., \cite{Rock}, Section 17) says that for any bounded set $\SS \subset \RR^r$, every point in the convex hull of $\SS$ can be expressed as a convex combination of at most $r+1$ points of $\SS$. Taking Lemma \ref{lMsm} into account, we obtain:
\begin{theorem}\label{t:s}
 For every design $\zeta \in \Upsilon$ there is a design $\zeta^s \in \Upsilon$ with the support size of at most $s$, such that $\M(\zeta)=\M(\zeta^s)$. In particular, there exists an optimal design supported on no more than $s$ points.
\end{theorem}
Theorem \ref{t:s} is a strengthening of the standard ``Carath\'{e}odory theorem'' used in optimal design (e.g., Theorem 2.2.3 in \citet{Fedorov} or Section 8.2. in \citet{Puk}): we obtain the standard result for $s=q$, but in general $s \leq q$.
\bigskip

Recall that $\zeta_\w \in \Upsilon$ is the design determined by the weights $\w \in \PP$ and $\vv_1,\ldots,\vv_\ell$ are the vertices of $\PP_*$. We will call the corresponding designs $\nu_j=\zeta_{\vv_j} \in \Upsilon_*$, $j=1,\ldots,\ell$, the ``vertex optimal designs'' (VODs). In simpler terms, for each vertex $\vv_j=(v_{j,1},\ldots,v_{j,d})'$ of $\PP_*$, $j \in \{1,\ldots,\ell\}$, the uniquely determined vertex optimal design is 
\begin{equation*} \nu_j =
\left[
\begin{array}{cccc}
\y_1 & \y_2 & \cdots & \y_d \\ \hline
v_{j,1} & v_{j,2} & \cdots & v_{j,d} \\
\end{array}
\right] \in \Upsilon_*.
\end{equation*}
Since the mapping $\w \mapsto \zeta_\w$ is a linear isomorphism between $\PP_*$ and $\Upsilon_*$, the role of the vertices $\vv_1,\ldots,\vv_\ell$ in $\PP_*$ is analogous to the role of the VODs $\nu_1,\ldots,\nu_\ell$ in $\Upsilon_*$.
\bigskip

As a consequence of Lemma \ref{lMsm} we see that the dimension of $\Upsilon_*$, understood as the dimension of its affine hull in $\Upsilon^{\pm}$, is $t=d-s$. The knowledge of $t$ and the inequality $s \geq m$ provides another variant of the Carath\'{e}odory theorem expressed in terms of the VODs:
\begin{theorem}\label{t:t}
    Any optimal design can be represented as a convex combination of $r$ VODs, where $r \leq t+1$.
\end{theorem}

The main theorem characterizing the VODs, based directly on Theorem \ref{tEquivPOW}, is

\begin{theorem}\label{tEquivPOD}
The following statements are equivalent:
\begin{enumerate}
    \item $\nu$ is a vertex optimal design, i.e., $\nu$ is the unique solution to the optimization problem $\textstyle \min \{\sum_{i=1}^d c_i\zeta(\y_i): \zeta \in \Upsilon_*\}$ for some coefficients $c_1,\ldots,c_d \in \RR$.
    \item $\nu$ is an extreme optimal design, i.e., $\nu \in \Upsilon_*$, and if $\zeta_1,\zeta_2 \in \Upsilon_*$ are such that $\nu = \alpha \zeta_1 + (1-\alpha)\zeta_2$ for $\alpha \in (0,1)$ then $\zeta_1=\zeta_2=\nu$. 
    \item $\nu$ is an optimal design with ``independent support information matrices'', i.e., $\nu \in \Upsilon_*$, and the matrices $\f(\y)\f'(\y)$, for $\y \in \supp(\nu)$, are linearly independent.
    \item $\nu$ is a minimal optimal design, i.e., $\nu \in \Upsilon_*$, and if $\zeta \in \Upsilon_*$ is such that $\supp(\zeta) \subseteq \supp(\nu)$, then $\supp(\zeta) = \supp(\nu)$.
    \item $\nu$ is a unique optimal design on $\supp(\nu)$, i.e., $\nu \in \Upsilon_*$, and if $\zeta \in \Upsilon_*$ is such that $\supp(\zeta) \subseteq \supp(\nu)$, then $\zeta = \nu$.
\end{enumerate}
\end{theorem}

Note that Theorem \ref{tEquivPOD}, specifically part $1 \Leftrightarrow 3$, implies that the support size of any VOD is between $m$ and $s$. Theorem \ref{tEquivPOD} also provides bounds on the number $\ell$ of the VODs in terms of the numbers $d$, $s$ and $t$; see Appendix \ref{sup:bounds}.

\begin{example}\label{ex:5}
Consider Model \eqref{eq:exam} from Examples \ref{ex:1}-\ref{ex:4}. The vertices of the polytope $\Upsilon_*$ of all optimal designs correspond to the design weights $\vv_1$ and $\vv_2$ derived in Example \ref{ex:4}. That is, there are two VODs $\nu_1=\zeta_{\vv_1}$ and $\nu_2=\zeta_{\vv_2}$, which are uniform on the sets $\Y_1$ and $\Y_2$ containing points $(-1,-1,-1)'$, $(+1,+1,-1)'$, $(+1,-1,+1)'$, $(-1,+1,+1)'$,  and $(+1,+1,+1)'$, $(-1,-1,+1)'$, $(-1,+1,-1)'$, $(+1,-1,-1)'$, respectively. The set $\Upsilon_*$ consists of designs $\alpha \nu_1 + (1-\alpha)\nu_2$, where $\alpha \in [0,1]$, which assign the weight $\alpha/4$ to all points of $\Y_1$ and the weight $(1-\alpha)/4$ to all points of $\Y_2$. One can easily see that the VODs $\nu_1$ and $\nu_2$ indeed satisfy all characterizations given by Theorem \ref{tEquivPOD}.
\end{example}

\subsection{Rational optimal design problems}\label{subsec:Rational}

In Examples \ref{ex:3} to \ref{ex:5}, we provided the complete characterization of the optimal design polytope via its vertices for Model \eqref{eq:exam}. However, such analytical derivations become tedious with increasing number of factors and more complex models.

Fortunately, the problems of finding all VODs for many standard models (including Model \eqref{eq:exam}) are ``rational'', that is, the polytope of optimal weights has a rational H-representation \eqref{eq:H}. More precisely, it means that the matrices $\M_*$ and $\f_i\f'_i$ ($i=1,\ldots,d$) have rational elements. These conditions are automatically satisfied if $\f_1, \ldots, \f_d$ have rational elements, and the weights of at least one optimal design are rational numbers.

For a rational optimal design problem, all VODs $\nu_1,\ldots,\nu_\ell$ have rational weights. Moreover, a rational optimal design problem allows for the use of error-free rational-arithmetic algorithms to construct the theoretical forms of all VODs. In Section \ref{sec:Examples}, we therefore employ computer-assisted construction of VODs that are (analytically) optimal, rather than the manual construction of such designs.

The rational form of the weights of all VODs means that they are \emph{exact} designs of some sizes $N_1,\ldots,N_\ell$. It is of theoretical interest that in the rational optimal design problems, any optimal approximate design is a convex combination of at most $t+1$ exact optimal designs; see Theorem \ref{t:t}. From the more practical point of view, the VODs can be combined to provide an entire pool of optimal designs that can be realized with a finite number of trials. In particular, for some $N_0$ we can achieve an exact design of any size $N \geq N_0$ that is a multiple of the greatest common divisor $\textrm{gcd}(N_1,\ldots,N_\ell)$, cf. \citet{NijenhuisWilf}. We illustrate this idea for all models analyzed in Section \ref{sec:Examples}.

Finally, note that for a rational design problem, its rank $s$ can also be calculated in a perfectly reliable way using the rational arithmetic, even without the need to calculate the VODs. The problem rank alone provides the dimension of the set of all information matrices on $\Y$ (Lemma \ref{lMsm}), an upper bound on the minimal support size (Theorem \ref{t:s}) and the dimension $t=d-s$ of the polytope $\Upsilon_*$ of optimal designs (Lemma \ref{lMsm}).

\subsection{Specific optimal designs}\label{subsec:Specific}

Suppose that the optimal design is not unique, and we have the complete list of all VODs, characterizing the set $\Upsilon_*$ of all optimal designs. Then we can select one or more optimal designs with special properties, to broaden the choice for the experimenter. 

\subsubsection{Optimal designs with small supports}\label{subsub:smallsupp}

Due to Theorem \ref{tEquivPOD}, the VODs exactly coincide with the minimal optimal designs. Since the set of VODs is finite, it is straightforward to determine all \emph{absolutely} minimal optimal designs, by selecting the VODs with the smallest possible size of the support.

Special optimal designs with small supports can also be obtained by maximizing a strictly convex function $\Psi$ over $\Upsilon_*$. Such a $\Psi$ can only attain its maximum at one or more of the extreme points of $\Upsilon_*$; by Theorem \ref{tEquivPOD}, these extreme points are exactly VODs. For instance, since the negative entropy 
\begin{equation*}
    \Psi_H(\zeta)=\sum_{\y \in \Y} \zeta(\y) \log_2(\zeta(\y))
\end{equation*}
is strictly convex on $\Upsilon_*$, we are able to identify all minimum-entropy optimal designs by selecting the VODs with the smallest entropy.

There are numerous potential advantages to using optimal designs with small supports, as opposed to those with large supports: they can better facilitate replicated observations, are simpler and more economical to execute, and offer a more compact representation.

Another advantage of approximate designs with small supports, such as the VODs, is that they are generally more appropriate for the ``rounding'' algorithms, which convert approximate designs to the exact designs. For instance, the popular Efficient Rounding (see \citet{PukelsheimRieder92}) is only meaningful if the approximate design to be rounded has a support smaller than the required size $N$ of the exact design. The size of the support also appears in the formula for the lower bound on the efficiency of the resulting exact design, in the sense that the smaller the support, the better the bound (Section 7 in \citet{PukelsheimRieder92}). 

Moreover, it is possible to ``round'' an optimal approximate design by applying an advanced mathematical programming method (e.g., \citet{SagnolHarman15} or \citet{Ahipasaoglu21}) at the support of an optimal approximate design, which generally provides better exact designs than Efficient Rounding. This approach can only be practically feasible if the associated optimization problem has a small number of variables, that is, if the support of the optimal approximate design is small. 

\subsubsection{Optimal designs with large supports}\label{subsub:largesupp}

The list of all VODs also provides the complete characterization of all optimal designs whose support is maximum possible, i.e., with support $\Y$. Recall that we call such designs maximal optimal designs. Every such design is of the form 
 $\alpha_1\nu_1+\cdots+\alpha_\ell\nu_\ell$, where $\alpha_j \geq 0$, $\sum_j \alpha_j = 1$,
\begin{equation*}
\cup_{j \: : \: \alpha_j > 0} \: \supp(\nu_j) = \Y,
\end{equation*}
and vice-versa. For instance, we can construct a maximal optimal design with $\alpha_j=1/\ell$ for all $j=1,\ldots,\ell$. In contrast to the minimally supported VODs, the set of all maximally-supported optimal designs corresponds to the relative interior of $\Upsilon_*$.

It is also often possible to select an optimal design with a large support by minimizing a convex function $\Psi$ over $\Upsilon_*$. This can be performed by means of appropriate mathematical programming solvers applied to the problem 
\begin{equation}\label{eq:optA}
\min \Psi(\zeta_\w), \text{ s.t. } \w \in \PP_*.
\end{equation}
For instance, using solvers for convex optimization on $\Psi_H$ under the appropriate linear constraints, we can obtain the unique maximum-entropy optimal design. 

While it is possible to solve \eqref{eq:optA} without the knowledge of the vertices of $\PP_*$, the vertices allow us to use two alternative approaches. First, applying the representation $\TT_*$, see \eqref{eq:isomorph}, we can solve the convex optimization problem
\begin{equation}\label{eq:optB}
\min \Psi(\zeta_{\w}), \w = \U\t+\w^*, \text{ s.t. } \U\t \geq -\w^*.
\end{equation}
Here, the variable $\t$ is $t$-dimensional. Second, we can use the fact that the vertices $\vv_1,\ldots,\vv_\ell$ of $\PP_*$ generate all optimal weights; therefore, we can solve the convex optimization problem
\begin{equation}\label{eq:optC}
\textstyle \min \Psi(\zeta_\w), \w = \sum_{i=1}^{\ell} a_i \vv_i, \text{ s.t. } \a \geq \0_\ell, \1'_\ell\a = 1,
\end{equation}
where the variable $\a$ is $\ell$-dimensional. For small $t$ or small $\ell$, these two restatements can be much simpler problems than \eqref{eq:optA}. For a demonstration, see Section \ref{subsec:main}.

Optimal designs with large supports can sometimes be preferable to those with small supports, because they offer a more comprehensive exploration of the design space and are more robust with respect to misspecifications of the statistical model. Additionally, they are sometimes mathematically simpler, exhibiting a higher degree of symmetry (cf. Section 5.1. in \citet{HFR2020REX}).

\subsubsection{Cost-efficient optimal designs}\label{subsub:cost}

Finally, consider a linear cost function $\Psi_c$ defined on $\Upsilon_*$:
\begin{equation*}
\Psi_c(\zeta)=\sum_{\y \in \Y} c(\y) \zeta(\y),
\end{equation*}
where $c(\y) \in \RR$ are the costs associated with the trial in $\y \in \Y$. While a $\Psi_c$-minimizing optimal design can be computed by solving a problem of linear programming, the list of all VODs provides much more: a straightforward and complete characterization of the set of \emph{all} optimal solutions. Indeed, let $\nu_{j_1},\ldots,\nu_{j_L}$ be the VODs that minimize $\Psi_c$. Then the set of all convex combinations of these designs is the set of all $\Psi_c$-minimizing optimal designs. Geometrically, the set of $\Psi_c$-minimizing optimal designs is always a face of $\Upsilon_*$ (of some dimension between $0$ and $t$). For an illustration, see Section \ref{ssMod2}. 

As a special case, we can directly obtain the range of the optimal weights at $\y \in \Y$, for instance as a means to explore the possible costs associated with the trials at $\y$. For $\y$, the optimal weights $\zeta^*(\y)$, $\zeta^* \in \Upsilon_*$, range in the (possibly degenerate) interval
\begin{equation*}
 W(\y) = \left[\min_{\nu \in \Upsilon_*} \nu(\y), \max_{\nu \in \Upsilon_*} \nu(\y)\right],
\end{equation*}
where $\nu \to \nu(\y)$ is linear on $\Upsilon_*$. Note that the interval $W(\y)$ is in fact a one-dimensional projection of $\PP_*$. Analogously, we can compute and visualize projections of $\PP_*$ of dimensions greater than $1$, which can provide complete information on the possible ranges of subvectors of optimal weights; see Section \ref{ss:Mod4} for illustrations of such projections.

\section{Examples}\label{sec:Examples}

In this section, we focus on optimal designs for several common regression models with $k$ factors. In each of these models, the uniform design on the maximum optimal support $\Y$ is a maximal optimal design (see Section \ref{subsec:mos}). The size $d$ of $\Y$ grows superpolynomially with $k$, while the number $m$ of parameters only grows polynomially. The standard Carath\'{e}odory theorem (cf. Theorem \ref{t:s}) implies that there always exists an optimal design with the support size of at most $m(m+1)/2$. Therefore, there exists some $k^*$ such that the optimal design is not unique for all $k \geq k^*$. 

Furthermore, each of the optimal design problems from this section is a rational problem (Section \ref{subsec:Rational}), as can be easily seen from the form of the model and the maximal optimal design. Therefore, we can use the rational arithmetic to identify the characteristics of the polytope of optimal designs, specifically its affine dimension $t$ and, most importantly, the sets of all VODs, cf. Sections \ref{subsec:PODW} and \ref{subsec:POD}.\footnote{We provide a description of the VODs for $k$ smaller than a limit determined by the capacity of our computational equipment. For some larger values of $k$, we were only able to compute the rank $s$ of the problem, which provides the dimension $t$ of the polytope of optimal designs by Lemma \ref{lMsm}.} 

For each model in this section, the minimum-entropy optimal designs (cf. Section \ref{subsub:smallsupp}) coincide with the absolutely minimal optimal designs, and the unique maximum-entropy optimal design (cf. Section \ref{subsub:largesupp}) is the uniform design on $\Y$. 

Our calculations were performed with the help of the statistical software \texttt{R}; in particular, the rational polytope vertices were obtained using the packages \citet{gmp} and \citet{rcdd}. We used a 64-bit Windows 11 system with an Intel Core i7-9750H processor operating at 2.60 GHz. The \texttt{R} codes and full detailed tables of all VODs from Section \ref{sec:Examples} are available upon request from the authors.

\subsection{The first-degree model on $[0,1]^k$ without a constant term}\label{ss:Mod1}

Consider the model 
\begin{eqnarray}
    E(Y_\x)&=&\beta_1 x_1 + \cdots + \beta_k x_k, \label{mod:SBW}\\
    \x&=&(x_1,\ldots,x_k)' \in \X =[0, 1]^k.\nonumber 
\end{eqnarray}
The case of a single factor ($k=1$) is trivial and for two factors ($k=2$), all $\Phi_p$-optimal designs are unique; see Section 8.6 in \citet{Puk}.

Let $k \geq 3$. For $j=1,\ldots,k$, let $\Vc(j)$ be the set of all vectors in $\{0,1\}^k$ with exactly $j$ components equal to $1$. The following claims are straightforward to verify via \eqref{eq:eqt21} and \eqref{eq:eqt22}; cf{.} \citet{Puk}, page 376:  If $k$ is odd, the maximum optimal support is $\Y_D=\Y_A=\Vc(k/2+1/2)$ for both $D$- and $A$-optimality. If $k$ is even, the maximum optimal support is $\Y_D=\Vc(k/2) \cup \Vc(k/2+1)$ for $D$-optimality and $\Y_A=\Vc(k/2)$ for $A$-optimality. For any $k$, the design uniform on $\Y_D$ is maximal $D$-optimal, and the design uniform on $\Y_A$ is maximal $A$-optimal.

Consequently, it is possible to apply rational arithmetic calculations to prove that for $k \leq 5$ the uniform optimal design on $\Y_D$ is the unique $D$-optimal design, and the uniform design on $\Y_A$ is the unique $A$-optimal design. 

For $k=6$ and the $D$-criterion, the dimension of $\Upsilon_*$ is $t=14$, and there are $\ell=150$ VODs. To describe the set of all VODs, consider two designs to be isomorphic, if they are the same up to a relabeling of the factors. Then, the VODs can be split into two classes of isomorphic designs. The first class consists of the $30$ VODs that are isomorphic to 
\begin{small}
\begin{equation}\label{eq:SBW6D1}
\left[
\begin{array}{ccccccc}
0 & 0 & 0 & 1 & 1 & 1 & 1 \\
0 & 1 & 1 & 0 & 0 & 1 & 1 \\
1 & 0 & 1 & 0 & 1 & 0 & 1 \\
1 & 1 & 0 & 1 & 0 & 0 & 1 \\
1 & 1 & 0 & 0 & 1 & 1 & 0 \\
1 & 0 & 1 & 1 & 0 & 1 & 0 \\ \hline
\frac{1}{7} & \frac{1}{7} & \frac{1}{7} & \frac{1}{7} & \frac{1}{7} & \frac{1}{7} & \frac{1}{7} \\
\end{array}
\right].
\end{equation}
\end{small}
All $30$ VODs in this class can be obtained by permuting the (first $6$) rows of the table. Note that permuting the columns corresponds to the same VOD. One of these VODs has been computed by integer quadratic programming as an exact $D$-optimal design of size $7K$, $K \in \NN$; see Table 1 in \citet{AQuA2}. For $k=6$, these $30$ VODs are the absolutely minimal $D$-optimal designs.

The second class of VODs for $D$-optimality and $k=6$ consists of the $120$ VODs isomorphic to the uniform design
\begin{small}
\begin{equation*}
\left[
\begin{array}{ccccccccccc}
0 & 0 & 0 & 0 & 0 & 0 & 0 & 0 & 0 & 1 & 1\\
0 & 0 & 0 & 1 & 1 & 1 & 1 & 1 & 1 & 0 & 0\\
1 & 1 & 1 & 0 & 0 & 0 & 1 & 1 & 1 & 0 & 0\\
1 & 1 & 1 & 0 & 1 & 1 & 0 & 0 & 1 & 0 & 1\\
0 & 1 & 1 & 1 & 1 & 1 & 0 & 1 & 0 & 1 & 0\\
1 & 0 & 1 & 1 & 0 & 1 & 1 & 1 & 0 & 1 & 1\\ \hline
\frac{1}{21} & \frac{1}{21} & \frac{1}{21} & \frac{1}{21} & \frac{1}{21} & \frac{1}{21} & \frac{1}{21} & \frac{1}{21} & \frac{1}{21} & \frac{1}{21} & \frac{1}{21}\\
\end{array}
\right.
\end{equation*}
\begin{equation}\label{eq:SBW6D2}
\left.
\begin{array}{cccccccccc}
1 & 1 & 1 & 1 & 1 & 1 & 1 & 1 & 1 & 1 \\
0 & 0 & 0 & 0 & 1 & 1 & 1 & 1 & 1 & 1 \\
0 & 1 & 1 & 1 & 0 & 0 & 0 & 1 & 1 & 1 \\
1 & 0 & 0 & 1 & 0 & 1 & 1 & 0 & 0 & 1 \\
1 & 0 & 1 & 1 & 1 & 0 & 0 & 0 & 1 & 0 \\
1 & 1 & 0 & 0 & 0 & 0 & 1 & 1 & 0 & 0 \\ \hline
\frac{1}{21} & \frac{1}{21} & \frac{1}{21} & \frac{1}{21} & \frac{1}{21} & \frac{1}{21} & \frac{1}{21} & \frac{1}{21} & \frac{1}{21} & \frac{1}{21} \\
\end{array}
\right].
\end{equation}
\end{small}

For $k=6$ and for the criterion of $A$-optimality, $\Upsilon_*$ has dimension $t=5$, and there are $\ell=12$ VODs, all isomorphic to 
\begin{small}
\begin{equation}\label{eq:SBW6A}
\left[
\begin{array}{cccccccccc}
0 & 0 & 0 & 0 & 0 & 1 & 1 & 1 & 1 & 1 \\
0 & 0 & 1 & 1 & 1 & 0 & 0 & 0 & 1 & 1 \\
1 & 1 & 0 & 0 & 1 & 0 & 0 & 1 & 0 & 1 \\
1 & 1 & 0 & 1 & 0 & 0 & 1 & 0 & 1 & 0 \\
0 & 1 & 1 & 1 & 0 & 1 & 0 & 1 & 0 & 0 \\
1 & 0 & 1 & 0 & 1 & 1 & 1 & 0 & 0 & 0 \\ \hline
\frac{1}{10} & \frac{1}{10} & \frac{1}{10} & \frac{1}{10} & \frac{1}{10} & \frac{1}{10} & \frac{1}{10} & \frac{1}{10} & \frac{1}{10} & \frac{1}{10} \\
\end{array}
\right].
\end{equation}
\end{small}

\begin{figure}
\centering
\includegraphics[width=0.3\textwidth]{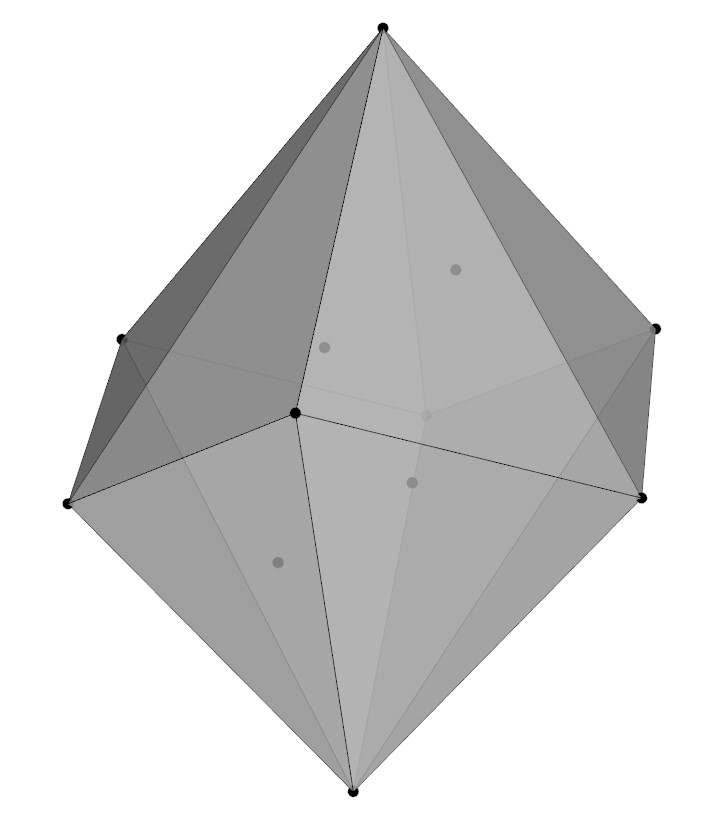}
\caption{A $3$-dimensional projection of the $5$-dimensional polytope $\PP_*$ of $A$-optimal weights for Model \eqref{mod:SBW} with $k=6$ factors. The projections of the $12$ vertices of $\PP_*$ are depicted by black dots.}
\label{fig:PPprojSBW}
\end{figure}

One of these VODs has been identified by integer quadratic programming as an $A$-optimal exact design of size $10K$, $K \in \NN$; see Table 2 in \citet{AQuA2}. For $k=6$, these $12$ VODs are the absolutely minimal $A$-optimal designs. A $3$-dimensional projection of $\PP_*$ and its $12$ vertices are depicted in Figure \ref{fig:PPprojSBW}.

For $k=7$, the sets of $D$- and $A$-optimal designs coincide. The dimension of $\Upsilon_*$ is $t=14$ and it has $\ell=150$ vertices; their structure is similar to the case of $D$-optimality for $k=6$.

For $k=8$ and the criterion of $D$ optimality, it holds that the size of the maximum optimal support is $d=126$, the problem rank is $s=36$, and the dimension of $\Upsilon_*$ is $t=90$. For $k=8$ and the criterion of $A$-optimality we have $d=70$, $s=38$ and $t=42$. We were not able to enumerate the individual VODs for this size.
\bigskip

\textbf{Relations to block designs.} 
For Model \eqref{mod:SBW}, a $k \times b$ matrix of support points of a design on $\Y_D$ or $\Y_A$ can be viewed as the incidence matrix of a binary block design with $k$ treatments and $b$ blocks. In this view, the support points of the $30$ VODs isomorphic to \eqref{eq:SBW6D1} are partially balanced designs (PBDs; see \citet{Wilson75}) with $r=4$, $\lambda=2$, and $b=7$. Here, $r$ is the common replication number, and $\lambda$ the common concurrence. Similarly, the $120$ VODs isomorphic to \eqref{eq:SBW6D2} are PBDs with $r=12$, $\lambda=6$, and $b=21$. In addition, the support points of \eqref{eq:SBW6A} correspond to a balanced incomplete block design (which is a PBD with a constant block size) for $6$ treatments, $r=5$, $\lambda=2$, $b=10$, and a common block size $3$ (cf. \citet{Puk}, p.\ 378, \citet{Cheng}).

\subsection{The first-degree model on $[-1,+1]^k$ without a constant term}
\label{ssMod2}

In this section we consider the model
 \begin{eqnarray}
    E(Y_\x)&=&\beta_1 x_1 + \cdots + \beta_k x_k, \label{mod:CBW}\\
    \x&=&(x_1,\ldots,x_k)' \in \X=[-1,+1]^k, \nonumber
\end{eqnarray}
where $k \geq 2$ (the case $k=1$ is trivial). For this model, it is simple to see that the design $\xi^*$ uniform on $\Y=\{-1,+1\}^k$ satisfies $\M(\xi^*)=\I_m$, and use conditions \eqref{eq:eqtSchur} and \eqref{eq:eqt22} to verify that $\xi^*$ is a maximal optimal design with respect to all $\Phi_p$, $p \in (-\infty,0]$; that is, the polytope of optimal designs does not depend on the choice of the Kiefer's criterion. The maximal optimal design is not uniquely optimal for any $k$. Clearly, the optimal design problem is rational for any $k$ and we can employ rational-arithmetic calculations to derive the following characterizations.

For $k=2$, the dimension of $\Upsilon_*$ is $t=2$, and there are $\ell=4$ VODs, each uniform on a two-point support. If we denote the elements of $\Y$ as $\y_1=(-1,-1)', \y_2=(-1,+1)', \y_3=(+1,-1)'$, and $\y_4=(+1,+1)'$, the VODs are uniform designs on the pairs $\{\y_1,\y_2\}$, $\{\y_1,\y_3\}$, $\{\y_2,\y_4\}$ and $\{\y_3,\y_4\}$. These designs are the absolutely minimal optimal designs. They can be combined to construct exact optimal designs of sizes $N=2K$, $K \in \NN$. 

For Model \eqref{mod:CBW}, we consider two designs to be isomorphic, if one can be obtained from the other one by any sequence of operations involving relabeling the factors, reverting the signs of any subset of the factors, and reverting all signs of the factors for any subset of the support points. For a design matrix representing a VOD, these operations mean permuting the rows, multiplying any set of rows by $-1$, and multiplying the factor levels of any set of columns by $-1$.

For $k=3$, the dimension of $\Upsilon_*$ is $t=4$, and there are $\ell=16$ VODs, all isomorphic to 
\begin{small}
\begin{equation*}
\left[
\begin{array}{cccc}
+1 & -1 & -1 & +1 \\
+1 & -1 & +1 & -1 \\
+1 & +1 & -1 & -1 \\ \hline
\frac{1}{4} & \frac{1}{4} & \frac{1}{4} & \frac{1}{4} \\
\end{array}
\right].
\end{equation*}
\end{small}

Note that the matrix of the support points is a submatrix of a $4 \times 4$ Hadamard matrix. These VODs are the absolutely minimal optimal designs. They can be combined to construct a variety of optimal designs that are exact of sizes $N=4K$, $K \in \NN$.

For $k=4$, the dimension of $\Upsilon_*$ is $t=9$, and there are $\ell=32$ VODs, all isomorphic to 
\begin{small}
\begin{equation*}
\left[
\begin{array}{cccc}
+1 & +1 & +1 & +1 \\
+1 & -1 & -1 & +1 \\
+1 & -1 & +1 & -1 \\
+1 & +1 & -1 & -1 \\  \hline
\frac{1}{4} & \frac{1}{4} & \frac{1}{4} & \frac{1}{4} \\
\end{array}
\right].
\end{equation*}
\end{small}

The support points form a $4 \times 4$ Hadamard matrix. In fact, the uniform design on the rows of any Hadamard matrix of order $4$ is a VOD and the support vectors of any VOD form a Hadamard matrix of order $4$. These $32$ VODs are the absolutely minimal optimal designs. They can be combined to construct exact optimal designs of sizes $N=4K$, $K \in \NN$.

For $k=5$, the dimension of $\Upsilon_*$ is $t=21$, and there are as many as $\ell=35328$ VODs. Of these, $2560$ are uniform designs on an $8$-point support (i.e., they are exact of size $N=8$) and $32678$ are nonuniform, supported on $11$ points ($N=12$). They can be used to construct exact optimal designs of sizes $N=4K$ for $K \in \NN$, $K \geq 2$. 

For $k=6$, the size of the maximum optimal support is $d=64$, the problem rank is $s=16$ and the dimension of $\Upsilon_*$ is $t=48$; for this size of the problem, the enumeration of the VODs was prohibitive for our computer equipment.
\bigskip

\textbf{Designs minimizing a linear cost.} As mentioned in Section \ref{subsub:cost}, the list of all VODs allows us to completely describe the class of linear-cost minimizing optimal designs. Consider Model \eqref{mod:CBW} with $k=5$ factors and assume that the cost of the trial in $\y \in \{-1,+1\}^5$ is $c=an_-+bn_+$, with any $a<b$, where $n_-$ is the number of levels $-1$ and $n_+$ is the number of levels $+1$ in $\y$. It is a matter of simple enumeration to determine that the VOD 
\begin{small}
\begin{equation}\label{eq:mincostA}
\left[
\begin{array}{cccccccc}
+1 & -1 & -1 & -1 & +1 & -1 & -1 & -1 \\
-1 & +1 & -1 & -1 & -1 & +1 & -1 & -1 \\
-1 & -1 & +1 & -1 & -1 & -1 & +1 & -1 \\
-1 & -1 & -1 & +1 & -1 & -1 & -1 & +1 \\
-1 & -1 & -1 & -1 & +1 & +1 & +1 & +1 \\ \hline
\frac{1}{8} & \frac{1}{8} & \frac{1}{8} & \frac{1}{8} & \frac{1}{8} & \frac{1}{8} & \frac{1}{8} & \frac{1}{8} \\
\end{array}
\right]
\end{equation}
\end{small}
minimizes this cost, as does any of the $4$ VODs that can be obtained from \eqref{eq:mincostA} by relabeling the factors. However, the VOD
\begin{small}
\begin{equation*}
\left[
\begin{array}{cccccc}
+1 & -1 & -1 & -1 & -1 & -1\\
-1 & +1 & -1 & -1 & -1 & +1\\
-1 & -1 & +1 & -1 & -1 & +1\\
-1 & -1 & -1 & +1 & -1 & -1\\
-1 & -1 & -1 & -1 & +1 & -1\\ \hline
\frac{1}{6} & \frac{1}{12} & \frac{1}{12} & \frac{1}{12} & \frac{1}{12} & \frac{1}{12}\\
\end{array}
\right.
\end{equation*}
\begin{equation}\label{eq:mincostB}
\left.
\begin{array}{ccccc}
-1 & -1 & -1 & -1 & -1 \\
+1 & -1 & +1 & -1 & -1 \\
-1 & +1 & -1 & +1 & -1 \\
+1 & +1 & -1 & -1 & +1 \\
-1 & -1 & +1 & +1 & +1 \\ \hline
\frac{1}{12} & \frac{1}{12} & \frac{1}{12} & \frac{1}{12} & \frac{1}{12} \\
\end{array}
\right]
\end{equation}
\end{small}
also minimizes this cost, as does any of the $4$ VODs that can be obtained from \eqref{eq:mincostB} by relabeling the factors. That is, the set of all cost-minimizing optimal designs is the set of all convex combinations of these $10$ VODs. Note that they provide exact optimal designs of sizes $N=4K$ for $K \in \NN$, $K \geq 2$.
\bigskip

In Sections \ref{subsec:main}, \ref{ss:Mod4}, \ref{ss:Mod5}, we consider two designs to be isomorphic, if one can be obtained from the other one by a sequence of relabeling the factors and reverting the signs of a subset of the factors.

\subsection{The first-degree model on $[-1,+1]^k$}\label{subsec:main}

Here we consider the model
 \begin{eqnarray}
    E(Y_\x)&=&\beta_1 + \beta_2 x_1 + \cdots + \beta_{k+1} x_{k}, \label{mod:MEM} \\
    \x&=&(x_1,\ldots,x_k)' \in \X=[-1,+1]^k, \nonumber
\end{eqnarray}
for $k \geq 1$. For this model, it can again be readily shown that the design $\xi^*$ uniform on $\Y=\{-1,+1\}^k$ has the information matrix $\I_m$, and it is possible to use conditions \eqref{eq:eqtSchur} and \eqref{eq:eqt22} to verify that $\xi^*$ is a maximal optimal design with respect to all Kiefer's criteria. Additionally, the optimal design problem itself is clearly rational.

The uniform design on $\Y$ is the unique optimal design if $k=1$ and $k=2$. For $k=3$, the dimension of $\Upsilon_*$ is $t=1$, and there are $\ell=2$ VODs, as described in Examples \ref{ex:1}-\ref{ex:5} for $\alpha=0$ and $\alpha=1$. Both these VODs are exact designs of size $N=4$, absolutely minimal optimal designs, and they can be combined to form exact optimal designs of sizes $N=4K$ for any $K \in \NN$.

The situation is more complex for $k=4$: the dimension of $\Upsilon_*$ is $t=5$, and there are $\ell=26$ VODs, which can be split into $3$ classes of isomorphic designs. The first subclass includes $8$ uniform designs isomorphic to 
\begin{small}
\begin{equation*}
\left[
\begin{array}{cccccccc}
+1 & +1 & +1 & +1 & -1 & -1 & -1 & -1 \\
+1 & -1 & -1 & +1 & +1 & -1 & -1 & +1 \\
+1 & -1 & +1 & -1 & +1 & -1 & +1 & -1 \\
+1 & +1 & -1 & -1 & +1 & +1 & -1 & -1 \\ \hline
\frac{1}{8} & \frac{1}{8} & \frac{1}{8} & \frac{1}{8} & \frac{1}{8} & \frac{1}{8} & \frac{1}{8} & \frac{1}{8} \\
\end{array}
\right].
\end{equation*}
\end{small}

This matrix of support points is formed by concatenating a normalized Hadamard matrix $\H_4$ of order $4$ and the matrix constructed from $\H_4$ by reverting the signs of the first row. Its transpose forms an $8 \times 4$ orthogonal array with two levels of strength $2$. 

The second class of VODs consists of two designs isomorphic to the uniform design
\begin{small}
\begin{equation*}
\left[
\begin{array}{cccccccc}
+1 & +1 & +1 & +1 & -1 & -1 & -1 & -1 \\
+1 & -1 & -1 & +1 & -1 & +1 & +1 & -1 \\
+1 & -1 & +1 & -1 & -1 & +1 & -1 & +1 \\
+1 & +1 & -1 & -1 & -1 & -1 & +1 & +1 \\ \hline
\frac{1}{8} & \frac{1}{8} & \frac{1}{8} & \frac{1}{8} & \frac{1}{8} & \frac{1}{8} & \frac{1}{8} & \frac{1}{8} \\
\end{array}
\right].
\end{equation*}
\end{small}

This support matrix can be constructed as $[\H_4, -\H_4]$. Its transpose forms an $8 \times 4$ orthogonal array of strength $3$. The VODs belonging to these two classes of uniform designs supported on $8$ points are the absolutely minimal optimal designs.

The third class of isomorphic VODs for $k=4$ contains $16$ designs, which are no longer uniform. They are all isomorphic to the following design supported on $11$ points: 
\begin{small}
\begin{equation*}
\left[
\begin{array}{cccccc}
-1 & -1 & -1 & +1 & -1 & -1\\
-1 & -1 & +1 & -1 & -1 & +1\\
-1 & +1 & -1 & -1 & +1 & -1\\
+1 & -1 & -1 & -1 & +1 & +1\\ \hline
\frac{1}{12} & \frac{1}{12} & \frac{1}{12} & \frac{1}{12} & \frac{1}{12} & \frac{1}{12}\\
\end{array}
\right.
\end{equation*}
\begin{equation*}
\left.
\begin{array}{ccccc}
-1 & +1 & +1 & +1 & +1 \\
+1 & -1 & -1 & +1 & +1 \\
+1 & -1 & +1 & -1 & +1 \\
-1 & +1 & -1 & -1 & +1 \\ \hline
\frac{1}{12} & \frac{1}{12} & \frac{1}{12} & \frac{1}{12} & \frac{1}{6} \\
\end{array}
\right].
\end{equation*}
\end{small}

If the point with weight $1/6$ were to be replicated twice, then the support points would form a transpose of a $12 \times 4$ orthogonal array with two levels of strength $2$. The $26$ VODs for $k=4$ provide exact designs of sizes $N=4K$, $K \in \NN$, $K \geq 2$.

For $k=5$, the dimension of $\Upsilon_*$ is $t=16$, and the number of VODs is $\ell=14110$. The support sizes of the VODs are $8$ (they are all exact designs of size $N=8$), $11$ ($N=12$), $12$ ($N=12$), $13$ ($N=20$), $15$ ($N=24$), and $16$ ($N=16,28,32,36$). The VODs can be combined to provide exact designs of sizes $N=4K$, $K \in \NN$, $K \geq 2$.

For $k=6$, the size of the maximum optimal support is $d=64$, the problem rank is $s=22$ and the dimension of $\Upsilon_*$ is $t=42$. The enumeration of the VODs for $k=6$ exceeded the capabilities of our computer equipment.
\bigskip

\textbf{Optimal designs minimizing a convex function}. To illustrate the search for an optimal design that minimizes a convex function (cf. Section \ref{subsub:largesupp}), consider Model \eqref{mod:MEM} with $k=4$ and the problem 
\begin{equation}\label{eq:riopt}
\min \Psi\left(\sum_{i=1}^d r_i \zeta(\y_i) \f_i\f'_i\right), \text{ s.t. } \zeta \in \Upsilon_*,
\end{equation}
where $r_1,\ldots,r_d \in (0,1]$ are given coefficients, and $\Psi=-\Phi_0$, i.e., the negative criterion of $D$-optimality. The problem \eqref{eq:riopt} can be interpreted as optimizing $\Phi_0$ on $\Upsilon_*$ in a heteroscedastic version of \eqref{mod:MEM}, with variances of the observations in $\y_1,\ldots,\y_d$ proportional to $1/r_1,\ldots,1/r_d$. An alternative interpretation is optimizing $\Phi_0$ on $\Upsilon_*$ in \eqref{eq:riopt} for proportions $1-r_1,\ldots,1-r_d$ of failed trials in $\y_1,\ldots,\y_d$.

For illustration, we chose the coefficients $r_i=1$ for $\y_i=(-1,-1,-1,-1)'$, $r_i=0.95$ for all four $\y_i$ that contain one level $+1$, $r_i=0.85$ for all six $\y_i$ that contain two levels $+1$, $r_i=0.70$ for all four $\y_i$ that contain three levels $+1$ and $r_i=0.50$ for $\y_i=(+1,+1,+1,+1)'$.

We formulated \eqref{eq:riopt} in the form \eqref{eq:optB} based on $\TT_*$. Instead of this form, one can also use the forms \eqref{eq:optA} or \eqref{eq:optC}, but our version works with $\t \in \RR^5$, whereas \eqref{eq:optA} is based on $\w \in \RR^{16}$ and \eqref{eq:optC} is based on $\a \in \RR^{26}$. Moreover, \eqref{eq:optB} requires maximizing a concave function over a feasible set with a guaranteed interior point $\0_5 \in \TT_*$. This allowed us to solve the problem with a function as simple as \texttt{constrOptim} of \texttt{R}. This procedure provided the design
\begin{small}
\begin{equation*}
\left[
\begin{array}{cccccc}
-1 & +1 & +1 & -1 & +1 & -1 \\
-1 & +1 & -1 & +1 & -1 & +1 \\
-1 & -1 & +1 & +1 & -1 & -1 \\
-1 & -1 & -1 & -1 & +1 & +1 \\ \hline
0.156 & 0.094 & 0.094 & 0.094 & 0.094 & 0.094\\
\end{array}
\right.
\end{equation*}
\begin{equation*}
\left.
\begin{array}{cccccc}
-1 & +1 & +1 & +1 & -1 & +1 \\
-1 & +1 & +1 & -1 & +1 & +1 \\
+1 & +1 & -1 & +1 & +1 & +1 \\
+1 & -1 & +1 & +1 & +1 & +1 \\ \hline
0.094 & 0.062 & 0.062 & 0.062 & 0.062 & 0.032 \\
\end{array}
\right].
\end{equation*}
\end{small}

\subsection{The first-degree model on $[-1,+1]^k$ with second-order interactions}\label{ss:Mod4}

Consider the model 
 \begin{eqnarray}
    E(Y_\x) &=& \hspace{-1mm} \beta_1 + \sum_{i=1}^{k} \beta_{i+1} x_i + \hspace{-4mm}\sum_{1 \leq i<j \leq k} \hspace{-3mm} \beta_{i,j} x_i x_j, \label{eq:modINT}\\
    \x &=& (x_1,\ldots,x_k)' \in \X=[-1,+1]^k, \nonumber
\end{eqnarray}
where $k \geq 2$ is the number of factors, and $m=1+k+\binom{k}{2}$ is the number of parameters. Model \eqref{eq:modINT} is again a rational model, and the design $\xi^*$ uniform on $\Y=\{-1,+1\}^k$ satisfying $\M(\xi^*)=\I_m$ is a maximal $\Phi_p$-optimal design for any $p \in (-\infty, 0]$, as can be confirmed by verifying conditions \eqref{eq:eqtSchur} and \eqref{eq:eqt22}. It is the unique optimal design for $k=2,3,4$. 

However, for $k=5$, an infinite number of optimal designs already exist; the dimension of $\Upsilon_*$ is $t=1$, and there are $\ell=2$ VODs, mutually isomorphic. One of these two VODs is the uniform design  
\begin{small}
\begin{equation*}
\left[
\begin{array}{cccccccc}
-1 & -1 & -1 & -1 & -1 & -1 & -1 & -1\\
-1 & -1 & -1 & -1 & +1 & +1 & +1 & +1\\
-1 & -1 & +1 & +1 & -1 & -1 & +1 & +1\\
-1 & +1 & -1 & +1 & -1 & +1 & -1 & +1\\
+1 & -1 & -1 & +1 & -1 & +1 & +1 & -1\\ \hline
\frac{1}{16} & \frac{1}{16} & \frac{1}{16} & \frac{1}{16} & \frac{1}{16} & \frac{1}{16} & \frac{1}{16} & \frac{1}{16} \\
\end{array}
\right.
\end{equation*}
\begin{equation*}
\left.
\begin{array}{cccccccc}
+1 & +1 & +1 & +1 & +1 & +1 & +1 & +1 \\
-1 & -1 & -1 & -1 & +1 & +1 & +1 & +1 \\
-1 & -1 & +1 & +1 & -1 & -1 & +1 & +1 \\
-1 & +1 & -1 & +1 & -1 & +1 & -1 & +1 \\
-1 & +1 & +1 & -1 & +1 & -1 & -1 & +1 \\ \hline
\frac{1}{16} & \frac{1}{16} & \frac{1}{16} & \frac{1}{16} & \frac{1}{16} & \frac{1}{16} & \frac{1}{16} & \frac{1}{16} \\
\end{array}
\right].
\end{equation*}
\end{small}

Its support points form the transpose of a $16 \times 5$ orthogonal array of strength $4$. Both of these VODs are absolutely minimal optimal designs. They can be combined to obtain exact optimal designs of sizes $N=16K$, $K \in \NN$.

\begin{figure*}
\centering
\includegraphics[width=0.9\textwidth]{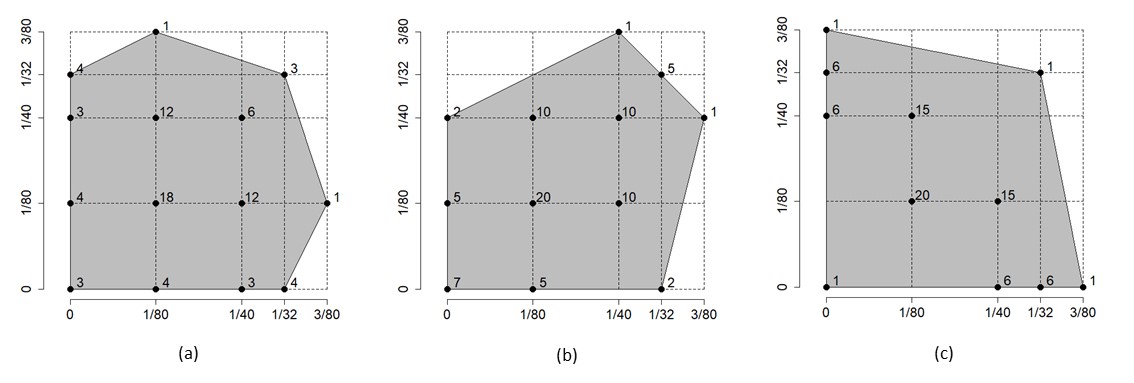}
\caption{Projections of the polytope $\PP_*$ for Model \eqref{eq:modINT} with $k=6$ factors to two-dimensional planes determined by various pairs of coordinate axes. More precisely, the panels (a), (b), and (c) depict the pairs $(\zeta^*(\y_{i_1}),\zeta^*(\y_{i_2}))$ for optimal designs $\zeta^* \in \Upsilon_*$, where $i_1, i_2 \in \{1,\ldots,d\}$ are such that $\y'_{i_1}\y_{i_2} \in \{-2,0\}$, $\y'_{i_1}\y_{i_2} \in \{-4,2\}$ and $\y'_{i_1}\y_{i_2} \in \{-6,4\}$, respectively. The projections of the $78$ vertices of $\PP_*$ are depicted by black dots. Note that the projections of groups of vertices coincide: the number near a dot signifies the number of vertices whose projections overlap.}
\label{fig:PPproj2D}
\end{figure*}

For $k=6$, the dimension of $\Upsilon_*$ is $t=7$ and there are $\ell=78$ VODs. The visualizations of selected $2$-dimensional projections of the corresponding $\PP_*$ are in Figure \ref{fig:PPproj2D}. Of these $78$ VODs, $14$ are uniform on a $32$-point support (exact of sizes $N=32$), and $64$ are nonuniform on a $57$-point support ($N=80$). Combining these VODs yields exact optimal designs of sizes $N=32$ and $N=16K$ for all $K \geq 4$.

The computations were prohibitive for Model \eqref{eq:modINT} with $k \geq 7$.

\subsection{The additive second-degree model on $[-1,+1]^k$}\label{ss:Mod5}

As the last example, we consider the model 
 \begin{eqnarray}
    E(Y_\x) &=& \beta_1 + \sum_{i=1}^{k} \beta_{i+1} x_i + \sum_{i=1}^{k} \beta_{i+1+k} x_i^2, \label{mod:QWOI} \\
    \x&=&(x_1,\ldots,x_k)' \in \X=[-1,+1]^k, \nonumber
\end{eqnarray}
where $k \geq 1$, and the number of parameters is $m=2k+1$. Model \eqref{mod:QWOI} is rational for $D$-optimality, as it is possible to verify that the uniform design on $\Y=\{-1,0,1\}^k$ is a maximal $D$-optimal\footnote{For this model, we consider only the criterion of $D$-optimality, since analytic formulas of other $\Phi_p$-optimal designs for Model \eqref{mod:QWOI} with $k>1$ do not seem to be available in the literature, nor trivial to derive.} design; cf. \citet{Puk} (Sections 9.5 and 9.9) for $k=1$, and \citet{SchwabeWierich} for $k>1$.  For $k=1$ and $k=2$, the uniform design on $\Y$ is the unique $D$-optimal design. 

However, for $k=3$ the dimension of the polytope of optimal designs is already $t=8$ and there are $\ell=66$ VODs. First, there are $12$ VODs, uniform on $9$ points, which form two distinct isomorphism classes. A representative of the $4$ members of the first class of VODs is the design
\begin{small}
\begin{equation}\label{des:D1}
\left[
\begin{array}{rrrrrrrrr}
-1 & +1 & 0 & +1 & 0 & -1 & 0 & -1 & +1 \\
-1 &  0 & +1 & -1 & 0 & +1 & -1 &  0 & +1 \\
-1 & -1 & -1 &  0 & 0 &  0 & +1 & +1 & +1 \\ \hline
\frac{1}{9} & \frac{1}{9} & \frac{1}{9} & \frac{1}{9} & \frac{1}{9} & \frac{1}{9} & \frac{1}{9} & \frac{1}{9} & \frac{1}{9} \\
\end{array}
\right].
\end{equation}
\end{small}

A representative of the $8$ members of the second class of VODs is the design
\begin{small}
\begin{equation}\label{des:D2}
\left[
\begin{array}{rrrrrrrrr}
-1 & +1 & 0 & 0 & -1 & +1 & +1 & 0 & -1 \\
-1 &  0 & +1 & -1 & 0 & +1 & -1 & 0 & +1 \\
-1 & -1 & -1 & 0 & 0 & 0 & +1 & +1 & +1 \\ \hline
\frac{1}{9} & \frac{1}{9} & \frac{1}{9} & \frac{1}{9} & \frac{1}{9} & \frac{1}{9} & \frac{1}{9} & \frac{1}{9} & \frac{1}{9} \\
\end{array}
\right].
\end{equation}
\end{small}

\begin{figure*}
\centering
\includegraphics[width=0.9\textwidth]{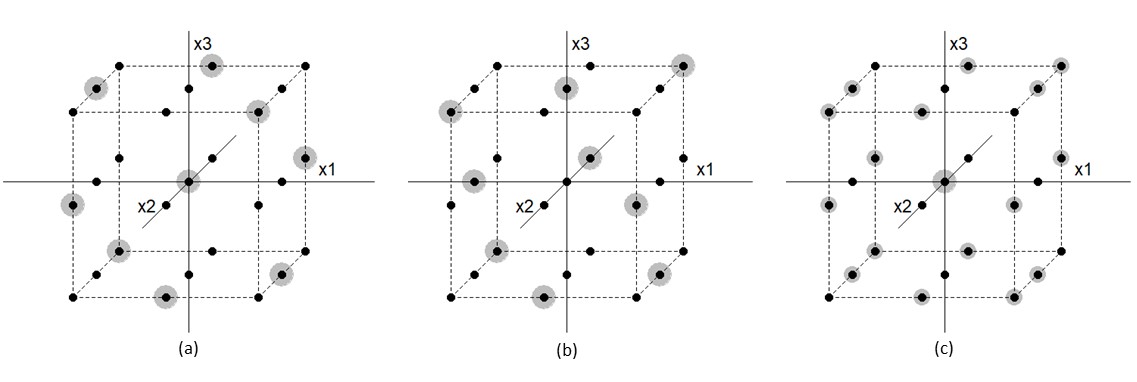}
\caption{Vertex $D$-optimal designs for Model \eqref{mod:QWOI} with $k=3$ factors. The panel (a) displays a representative of the class of $4$ isomorphic VODs uniform on a $9$-point support. More precisely, the gray circles depict the support points of the VOD \eqref{des:D1}. Similarly, the panel (b) displays the representative \eqref{des:D2} of the class of $8$ isomorphic VODs uniform on a $9$-point support, and the panel (c) displays the design \eqref{des:D3} which is one of the remaining $54$ non-uniform VODs supported on $17$ points. The maximum optimal support $\{-1,0,+1\}^3$ is visualized by black dots, and the edges of the cube $[-1,+1]^3$ are visualized by dashed lines.}
\label{fig:VODs}
\end{figure*}

The support points of these $12$ VODs form the transpose of an orthogonal array of strength $2$ with $v=3$ levels and $k=3$ factors. The nice geometric properties of these designs are apparent from Figure \ref{fig:VODs}. These VODs are absolutely minimal optimal designs.

The remaining $54$ VODs are all designs supported on $17$ design points from $\Y$, such that $16$ of these points have weight $1/18$, one point has weight $1/9$, and the projections to all three two-dimensional sub-spaces of these designs are uniform on the $3 \times 3$ grid. They form $5$ distinct isomorphism classes which we will not detail. One of these VODs is
\begin{small}
\begin{equation*}
\left[
\begin{array}{rrrrrrrrr}
-1 & -1 & -1 & -1 & -1 & -1 &  0 &  0 &  0 \\
-1 & -1 &  0 &  0 & +1 & +1 & -1 & -1 &  0\\
-1 &  0 & -1 & +1 &  0 & +1 & -1 & +1 &  0 \\ \hline
\frac{1}{18} & \frac{1}{18} & \frac{1}{18} & \frac{1}{18} & \frac{1}{18} & \frac{1}{18} & \frac{1}{18} & \frac{1}{18} & \frac{1}{9}\\
\end{array}
\right.
\end{equation*}
\begin{equation}\label{des:D3}
\left.
\begin{array}{rrrrrrrr}
  0 &  0 & +1 & +1 & +1 & +1 & +1 & +1 \\
 +1 & +1 & -1 & -1 &  0 &  0 & +1 & +1 \\
 -1 & +1 &  0 & +1 & -1 & +1 & -1 &  0 \\ \hline
\frac{1}{18} & \frac{1}{18} & \frac{1}{18} & \frac{1}{18} & \frac{1}{18} & \frac{1}{18} & \frac{1}{18} & \frac{1}{18} \\
\end{array}
\right].
\end{equation}
\end{small}

If the trial at $(0,0,0)'$ were replicated twice, the matrix of the support points would form the transpose of an $18 \times 3$ orthogonal array with $3$ levels of strength $2$. The $66$ VODs for $k=3$ can be combined to obtain various exact optimal designs of sizes $N=9K$, $K \in \NN$.

As noted by one reviewer, Model \eqref{mod:QWOI} on $\Y$ can be linearly reparametrized as a main-effects ANOVA model with $k$ factors on $3$ levels each. Consequently, the $D$-optimal designs in these two models should coincide. Indeed, it is possible to verify that the 3-dimensional projections of the widely used Taguchi L18 design for the ANOVA model form (some) $D$-optimal designs for Model \eqref{mod:QWOI} with $k=3$ factors. These $D$-optimal designs can be obtained as $\frac{1}{2}\nu_a+\frac{1}{2}\nu_b$, where $\nu_a$ and $\nu_b$ are VODs from the two isomorphism classes represented by designs \eqref{des:D1} and \eqref{des:D2}.

For $k=4$, the size of the maximum optimal support is $d=81$, the problem rank is $s=33$, and the dimension of the polytope of optimal designs is $t=48$. However, for $k=4$ the enumeration of the VODs was too time-consuming.

\section{Conclusions}\label{sec:Further}

We brought attention to the fact that the set of all optimal approximate designs is generally a polytope. We have shown that its characteristics can be beneficial to both the theory and applications of experimental design. Particularly noteworthy are the vertices of this polytope -- the vertex optimal designs -- because of their several unique properties, especially their small supports, and the potential to generate all optimal designs.

Another key point is the application of a rational-arithmetic method that accurately identifies all vertex optimal designs for rational optimal design problems. The use of rational vertex enumeration methods is a new tool of constructing optimal approximate as well as optimal exact designs in an error-free manner, complementing the plethora of analytical techniques in the literature.

Importantly, we have obtained the vertex optimal designs for several commonly used models, thus providing complete information on the optimal approximate designs in these cases. We have demonstrated that in the studied cases, these designs, along with their combinations, correspond to exact optimal designs of various sizes, often related to block designs or orthogonal arrays.

To the best of our knowledge, this paper presents the first systematic approach to studying the set of all optimal designs, and it motivates a variety of intriguing topics for further research. For instance, it is possible to investigate the polytope of optimal designs for important non-rational design problems (including the optimal design problems for the full quadratic model with many factors) and less common statistical models, such as models for paired comparisons (e.g., \cite{Grasshoff04}) or models with restricted design spaces (e.g., \cite{Freise}). One can also explore extending the results to more specialized optimality criteria, such as those focusing on the estimation of a subset of parameters (see, for instance, Sections 10.2 to 10.5 in \citet{AtkinsonEA07}). In addition, a general avenue of future research is the application of rational-arithmetic calculations in other problems of experimental design.

\backmatter

\bmhead{Acknowledgements}
We are grateful to Professor Rainer Schwabe from Otto von Guericke University Magdeburg and Associate Professor Martin M\'a\v{c}aj from Comenius University Bratislava for helpful discussions on some aspects of the paper. We would also like to thank the anonymous referees for their insightful comments that helped improve the paper.

\section*{Declarations}

\begin{itemize}
\item Funding. This work was supported by the Slovak Scientific Grant Agency (VEGA), Grant 1/0362/22.
\item Competing interests. The authors have no relevant financial or non-financial interests to disclose.
\item Ethics approval and consent to participate. Not applicable.
\item Consent for publication. Not applicable.
\item Data availability. Not applicable.
\item Materials availability. Not applicable.
\item Code availability. All codes used for the numerical results are available upon request from the authors.
\item Author contribution. All coauthors contributed to the conceptualization, methodology, mathematical proofs, software, writing and reviewing of the article.
\end{itemize}


\begin{appendices}

\section{Notation}\label{sup:notation}

Below is an overview of the notation used throughout the paper and the appendices.

\begin{itemize}
    \renewcommand\labelitemi{}
    \item $\mathbf{0}_d\ldots$ $d\times 1$ vector of all zeros
    \item $\mathbf{1}_d\ldots$ $d\times 1$ vector of all ones
    \item $a_1,\ldots,a_d\ldots$ components of a vector $\a \in \RR^d$
    \item $\I_m\ldots$ $m\times m$ identity matrix
    \item $\mathcal{C}(\A)\ldots$ column space of the matrix $\A$
    \item $\mathcal{N}(\A)\ldots$ null space of the matrix $\A$
    \item $\mathrm{tr}(\A)\ldots$ trace of the matrix $\A$
    \item $\mathrm{det}(\A)\ldots$ determinant of the matrix $\A$
    \item $\mathrm{vech}(\A)\ldots$ half-vectorization the matrix $\A$
    \item $\a', \A'\ldots$ transpose of $\a,\A$, respectively
    \item $Y_x = \f'(\x)\beta + \epsilon \ldots$ linear model
    \item $\f \in \RR^m \ldots$ regression function
    \item $m \ldots$ number of model parameters in $\beta$
    \item $N \ldots$ number of trials
    \item $\x \ldots$ design point
    \item $\X \ldots$ set of all design points, design space
    \item $\xi \ldots$ (approximate) design
    \item $\supp(\xi) \ldots$ support of $\xi$
    \item $\M(\xi) \ldots$ information matrix of $\xi$
    \item $\Phi_p \ldots$ Kiefer's $\Phi_p$-criterion
    \item $\xi^*, \zeta^* \ldots$ optimal design
    \item $\M_* \ldots$ optimal information matrix
    \item $\Y \ldots$ maximum optimal support
    \item $d \ldots$ size of $\Y$
    \item $\y_1, \ldots, \y_d \ldots$ elements of $\Y$
    \item $\f_i = \f(\y_i)$
    \item $\zeta \ldots$ design supported on $\Y$
    \item $\Upsilon \ldots$ set of all designs supported on $\Y$
    \item $\Upsilon_* \ldots$ set of all optimal designs
    \item $\w \in \RR^d \ldots$ vector of design weights
    \item $\supp(\w) \ldots$ support of $\w$
    \item $\PP \ldots$ simplex of the weights $\w$ of all designs on $\Y$
    \item $\PP_* \ldots$ polytope of all optimal weights
    \item $\w_\zeta \ldots$ vector of weights corresponding to $\zeta$
    \item $\zeta_\w \ldots$ design on $\Y$ with design weights $\w$
    \item $\A=\bigl[\vech\left(\f_1\f'_1\right), \ldots, \vech\left(\f_d\f'_d\right)\bigr]$
    \item $q = m(m+1)/2$ (number of rows of $\A$)
    \item $s = \mathrm{rank}(\A)$ (problem rank)
    \item $t = d - s$ (dimension of $\PP_*$)
    \item $\mathbb{T}_* \ldots$ polytope in $\RR^t$ that is isomorphic to $\PP_*$
    \item $\ell \ldots$ number of vertices of $\PP_*$, number of VODs
    \item $\vv_1, \ldots, \vv_\ell \ldots$ vertices of $\PP_*$
    \item $\nu_1, \ldots, \nu_\ell \ldots$ vertext optimal designs (VODs)
    \item $k \ldots$ number of factors
    \item $\mathcal{V}(j) \ldots$ vectors in $\{0,1\}^k$ with $j$ ones
    \item $\Y_D \ldots$ set $\Y$ for $D$-optimality
    \item $\Y_A \ldots$ set $\Y$ for $A$-optimality
    \item $r, \lambda, b \ldots$ replication number, concurrence number, number of blocks of a partially balanced design
    \item $\H_m \ldots$ Hadamard matrix of order $m$
\end{itemize}

\section{Proofs}\label{sup:proofs}%

\subsection{Proof of Lemma \ref{lPolyShort}.}

\begin{proof}
Let $\w \geq \0_d$ satisfy $\M(\zeta_\w)=\M_*$. Then $\tr(\M_*^{p})=\tr(\M(\zeta_\w)\M_*^{p-1})=\sum_{i=1}^d w_i \f'_i\M_*^{p-1}\f_i
=\tr(\M_*^{p})(\1'_d\w)$,
where the second equality follows form the basic properties of the trace, and the last equality follows from (2), i.e., from the equivalence theorem $\f'(\x)\mathbf{M}^{p-1}_*\f(\x)=\tr(\mathbf{M}^p_*)$ for all $\x \in \X$. As $\tr(\M_*^{p})>0$, we see that $\1'_d\w=1$, i.e., $\w \in \PP$.
\end{proof}

\subsection{Proof of Lemma \ref{lMsm}.}

\begin{proof}

    (a): Because $\A$ is a $q \times d$ matrix, the second inequality is clear. To prove the first inequality, observe that since $\M_*$ is nonsingular,  $\f_1,\ldots,\f_d$ span $\mathbb{R}^m$; i.e., there exist $m$ linearly independent $\f_i$'s. Without loss of generality assume that $\f_1, \ldots, \f_m$ are linearly independent. We shall prove that the matrices $\f_1\f_1', \ldots, \f_m\f_m'$ are then also linearly independent (in the space of all $m \times m$ matrices):
    
    Let $\sum_i\alpha_i \f_i \f_i' = \0$ for some $\alpha_1, \ldots, \alpha_m$. This can be expressed as $\B\D\B' = \0$, where
    $\B = [\f_1, \ldots, \f_m]$, and $\D$ is the diagonal matrix with $\alpha_1, \ldots, \alpha_m$ on the diagonal.
    
    Because $\f_1, \ldots, \f_m$ are linearly independent, $\B$ is an $m \times m$ non-singular matrix. Therefore $\rank(\D) = \rank(\0) = 0$, where we used the fact that the multiplication by a non-singular matrix preserves rank. It follows that $\D=\0$; i.e., $\alpha_1 = \ldots = \alpha_m = 0$, which proves the linear independence of $\f_1\f_1', \ldots \f_m\f_m'$. 
    
    Since $\vech$ is a linear isomorphism, it follows that $\vech(\f_1\f_1'), \ldots, \vech(\f_m\f_m')$ are linearly independent, which implies $\rank(\A) \geq m$.
    \bigskip

(b): First note that (b) trivially holds when $d=1$: then $s=1$ and the set $\{\M(\w): \w \in \PP\}$ contains just one matrix, so its dimension is $0 = s-1$. Suppose now that $d \geq 2$. Let  $\g_i := \vech(\f_i\f_i')$, $i=1,\ldots, d$. The $\vech$ function is a linear isomorphism between $\mathbb{S}^m$ and $\mathbb{R}^q$, so the dimension of $\{\M(\w): \w \in \PP\}$ is equal to the dimension of $\{\vech(\M(\w)): \w \in \PP\}$. The latter set is exactly the convex hull of $\g_1, \ldots, \g_d$, whose dimension is equal to the dimension of the affine hull of $\g_1, \ldots, \g_d$; i.e., the dimension of $\mathbb{M}:= \{\A\w: \1'_d\w=1\}$. Observe that the linearity of vech gives $\A\w = \vech(\sum_i w_i \f_i \f_i')$ and $\A\1_d = \vech(\sum_i \f_i \f_i')$. 

We will show that $\sum_i \f_i \f_i' \neq \sum_i w_i \f_i \f_i'$ for any vector $\w \in \mathbb{R}^d$ satisfying $\1_d'\w = 1$ (even for $\w$ with some negative components). If we denote $\H:=\M_*^{p-1}$ and $c:=\tr(\M_*^p$), the fact that $\f'(\x)\mathbf{M}^{p-1}(\tilde{\xi}^*)\f(\x)=\tr(\mathbf{M}^p(\tilde{\xi}^*))$  for all $\x \in \mathcal{S}(\tilde{\xi}^*)$ implies $\f_i'\H\f_i = c$ for all $i$. For contradiction, suppose that there exists $\w$ such that $\1_d'\w = 1$ and $\sum_i \f_i \f_i' = \sum_i w_i \f_i \f_i'$. It follows that $\tr(\H\sum_i \f_i \f_i') = \tr(\H\sum_i w_i \f_i \f_i')$. Properties of $\tr$ then yield $\sum_i \f_i' \H \f_i = \sum_i w_i \f_i' \H \f_i$; i.e., $cd = c \sum_i w_i=c$. This is, however, a contradiction, because $d \geq 2$.

As $\sum_i \f_i \f_i' \neq \sum_i w_i \f_i \f_i'$ for any $\w$ satisfying $\1_d'\w = 1$, we have $\A\1_d  \neq \A\w$ for any such $\w$. 
It follows that $\mathbb{M}$ is a strict subset of $\{\A\w: \w \in \RR^d\}$, which is $\mathcal{C}(\A)$. Therefore, the dimension of $\mathbb{M}$ must be less than the dimension of the column space $\mathcal{C}(\A)$, which is $s$. However, we can also write $\mathcal{C}(\A)$ as $\{\A(\w+\gamma \1_d): \1_d'\w=1, \gamma \in \mathbb{R}\} = \mathbb{M} + \{\gamma \A\1_d : \gamma \in \mathbb{R}\}$ (in the sense of the usual definition of the sum of two sets), which means that the dimension of $\mathbb{M}$ is at least $s-1$. Thus, the dimension of $\mathbb{M}$ is exactly $s-1$.

\bigskip
(c): Since the vector of weights $\widetilde{\w}^*$ that corresponds to the maximal optimal design satisfies $\widetilde{\w}^*>\0_d$ and $\A\widetilde{\w}^*=\vech(\M_*)$, the dimension of $\PP_*$ is the dimension $t$ of the null-space $\mathcal{N}(\A) \subseteq \RR^d$. From the rank-nullity theorem we therefore have that the dimension of $\PP_*$ is $d - \rank(\A) = d - s$.
\end{proof}

\subsection{Proof of Theorem \ref{tEquivPOW}.}\label{subap:proofPOW}

Let $\b = \vech(\M_*)$. Then, $\PP_*$ can be expressed as
$$
\PP_* = \{ \w \in \RR^d : \A\w = \b, \w \geq \0_d \}.
$$
Now some notation used in the theory of linear programming needs to be introduced:
\begin{itemize}
    \item A constraint is any row in $\A\w = \b$ and any inequality $w_i \geq 0$; i.e., $\PP_*$ is defined by $q+d$ constraints, where $q = m(m+1)/2$ is the number of rows of $\A$.
    \item A constraint is active at $\vv \in \mathbb{R}^d$ if it is satisfied as equality (that is, for $\vv \in \PP_*$ the active constraints are all the constraints in $\A\w = \b$, and all those constraints $w_i \geq 0$, for which $v_i = 0$). 
    \item The vector $\vv \in \mathbb{R}^d$ is a basic solution if $\A\vv = \b$, and there are $d$ linearly independent constraints among all the constraints that are active at $\vv$.
    \item $\vv$ is a basic feasible solution if it is a basic solution and $\vv \in \PP_*$ (i.e., if it also satisfies $\vv \geq \0_d$).
\end{itemize}

\begin{proof}
    Theorem 2.3 by \citet{BertsimasTsitsiklis} directly yields the equivalence of 1. and 2. That theorem also states that these are equivalent with $\vv$ being a basic feasible solution, and we shall show that this condition is equivalent to 4. and 5.
    
    Let $\vv$ be a basic feasible solution, which means that there are $d$ linearly independent active constraints at $\vv$. Let $I_0$ be the set of indices such that $v_i = 0$ for $i \in I_0$. Then Theorem 2.2 by \citet{BertsimasTsitsiklis} says that $\vv$ is the only vector that satisfies $\A\w = \b$ and $w_i = 0$ for $i \in I_0$. Now let $\z \in \PP_*$ satisfy $\supp(\z) \subseteq \supp(\vv)$, that is $z_i = 0$ for $i \in I_0$. It follows that $\z = \vv$ because the solution to $\A\w = \b$ and $w_i = 0$ for $i \in I_0$ was unique; thus showing 4. Note that this also proves that if $\vv$ is a basic feasible solution, then it is the unique design supported on $\supp(\vv)$, i.e., it proves 5.

    Conversely, let $\vv$ satisfy 4. and again denote $I_0$ as the set of indices for which $v_i = 0$. Now suppose that $\vv$ is not a basic solution, which means (in light of Theorem 2.2 by \citet{BertsimasTsitsiklis}) that there exists $\z \neq \vv$ such that $\A\z = \b$ and $z_i = 0$ for $i \in I_0$ (note that $z_j \geq 0$ need not be satisfied outside $I_0$). However, any point $\widetilde{\w}$ that lies on the line going through $\z$ and $\vv$ also satisfies $\A\widetilde{\w} = \b$, $\widetilde{w}_i = 0$ for $i \in I_0$. Now take $\widetilde{\w}$ as the point on this line that satisfies $\widetilde{w}_i \geq 0$ for $i \not\in I_0$ and such that $\widetilde{w}_j = 0$ for some $j \not\in I_0$. Such $\widetilde{\w}$ can be constructed as the intersection of the $\z\vv$ line and the relative boundary of $\PP_*$, and it exists because $\vv \in \PP_*$ and $v_i > 0$ for $i \not\in I_0$. This $\widetilde{\w} \in \PP_*$ satisfies $\supp(\widetilde{\w}) \subset \supp(\vv)$, which is in contradiction with 4. Note that if we assume 5. instead of 4., then $\widetilde{\w} \in \PP_*$ constructed above is also in contradiction with 5. because it satisfies $\supp(\widetilde{\w}) \subseteq \supp(\vv)$, but $\widetilde{\w} \neq \vv$. This concludes the proof that $\vv$ is a basic feasible solution if and only if it satisfies either 4. or 5. Now Theorem 2.3 by \citet{BertsimasTsitsiklis} gives the equivalence of 1., 4. and 5.

    If $\rank(\A) = q$, then Theorem 2.4 by \citet{BertsimasTsitsiklis} yields that $\vv$ is a basic feasible solution if and only if it satisfies 3. However, the proof of that theorem can be easily extended to the case of $\A$ of a general rank $s \leq q$. It follows that 3. is equivalent to 4. and 5.
\end{proof}

\section{Ad Example \ref{ex:4}}\label{sup:matrix}%

In Example 4, the full form of $\A\w = \vech(\I_4)$ is
\begin{small}
    \begin{equation}\label{eq:Awb}
    \begin{bmatrix}
        +1 & +1 & +1 & +1 & +1 & +1 & +1 & +1 \\
        -1 & -1 & -1 & -1 & +1 & +1 & +1 & +1 \\
        +1 & +1 & +1 & +1 & +1 & +1 & +1 & +1 \\
        -1 & -1 & +1 & +1 & -1 & -1 & +1 & +1 \\
        +1 & +1 & -1 & -1 & -1 & -1 & +1 & +1 \\
        +1 & +1 & +1 & +1 & +1 & +1 & +1 & +1 \\
        -1 & +1 & -1 & +1 & -1 & +1 & -1 & +1 \\
        +1 & -1 & +1 & -1 & -1 & +1 & -1 & +1 \\
        +1 & -1 & -1 & +1 & +1 & -1 & -1 & +1 \\
        +1 & +1 & +1 & +1 & +1 & +1 & +1 & +1
    \end{bmatrix} \w =
    \begin{bmatrix}
        1 \\ 0 \\ 1 \\ 0 \\ 0 \\ 1 \\ 0 \\ 0 \\ 0 \\ 1
    \end{bmatrix}.
    \end{equation}
\end{small}

\section{Vertex bounds}\label{sup:bounds}%

If $\nu_1,\nu_2$ are two distinct vertex optimal designs (VODs), it follows that neither set $\supp(\nu_1)$ nor $\supp(\nu_2)$ can be a subset of the other, which follows from the part $1 \Leftrightarrow 5$ of Theorem \ref{tEquivPOD}. That is, the system of the supports of the VODs forms a Sperner family; this provides a simple upper limit on the number of the VODs via the Sperner's theorem in terms of the simplest characteristic $d$: $\ell \leq \binom{d}{\lfloor d/2 \rfloor}$. If we also know the problem rank $s$ or the dimension $t$ of $\Upsilon_*$, we can improve the Sperner's bound as follows: since each $s$-touple of independent elementary information matrices ``supports'' at most one VOD (see $1 \Leftrightarrow 3$ in Theorem 2), we have $\ell \leq \binom{d}{s}=\binom{d}{t}$. For many situations, we can further improve the bound by using the H-representation of the full-dimensional isomporphic polytope $\mathbb{T}_*$ and the McMullen’s Upper Bound Theorem (see, e.g., the introduction in \citet{avis-devroye}), which gives 
\begin{equation*}
    \ell \leq \binom{d-\lfloor \frac{t+1}{2} \rfloor}{s} + \binom{d-\lfloor \frac{t+2}{2} \rfloor}{s}.
\end{equation*}

On the other hand, since the support of each VOD must be at most $s$, and any maximal optimal design must be a convex combination of some VODs, we must have $\ell \geq \lceil d/s \rceil$. Also, since the dimension of $\Upsilon_*$ is $t$ and the VODs generate $\Upsilon_*$, we must have $\ell \geq t+1$. 

Specifically, for $t=0$ ($t=1$, $t=2$) we have $\ell=1$ ($\ell=2$, $\ell \in \{3,\ldots,d\}$).
    
\end{appendices}


\end{document}